\documentclass[preprint,12pt,3p]{elsarticle}
\usepackage{tabularx}   
\usepackage{booktabs}   
\renewcommand{\arraystretch}{1.1}  
\usepackage{multirow}
\usepackage{float}
\usepackage{color}
\usepackage{soul}
\usepackage{placeins}
\usepackage{graphicx}
\usepackage{subcaption}
\usepackage[percent]{overpic}
\usepackage{url}
\usepackage{hyperref}
\usepackage{algorithm}
\usepackage{algpseudocode}

\usepackage{amssymb}

\usepackage{amsmath}
\usepackage{yfonts}

\journal{ }

\begin{document}

\begin{frontmatter}


\title{Data-Free Weak-Form Staggered Neural Operators for Magneto-Mechanical Coupling in Finite-Strain Elastomers}

\author[label0]{Alireza Yazdandousthamedani}
\address[label0]{Institute for Structural Analysis, Technische Universität Dresden, 01062 Dresden, Germany}

\author[label0]{Ahmad Moeineddin}

\author[label1]{Reza Najian Asl}
\address[label1]{Chair of Structural Analysis, Technical University of Munich, 80333 München, Germany}

\author[label2]{Shahed Rezaei\corref{cor1}}
\address[label2]{ACCESS e.V., 52062 Aachen, Germany}
\ead{s.rezaei@access-technology.de}

\author[label0]{Michael Kaliske\corref{cor1}}
\ead{michael.kaliske@tu-dresden.de}

\cortext[cor1]{Corresponding author}

\begin{abstract}

Magneto-active elastomers, as a class of smart materials, exhibit strongly coupled magnetic and mechanical behavior at finite strains. Considering variations in microstructure, material properties, and geometry can lead to computationally expensive analyses. Building on the finite operator learning (FOL) framework, this study develops a data-free, physics-informed operator-learning framework for families of coupled finite-strain magneto-mechanical boundary-value problems. The governing magnetostatic and mechanical equations are enforced through finite-element weak-form residuals, allowing the neural operators to be trained without labeled finite-element solution data. The main contribution of this study is the development of a weak-form staggered neural operator (WSNO) framework for strongly coupled magneto-mechanical saddle-point problems. Separate magnetic and mechanical neural operators are trained alternately using a staggered optimization strategy, while their physical coupling is retained through the constitutive relations and residual evaluations. The resulting framework learns mappings from parameterized material and geometric descriptions to the corresponding coupled magnetic and mechanical solution fields. The proposed framework is investigated across several settings, including heterogeneous random-inclusion microstructures, varying magnetic phase contrast, area-fraction-dependent geometries, strongly out-of-distribution material morphologies, and three-dimensional geometry-parametric problems. In addition, the learned operator is combined with a neural-initialized Newton strategy, in which the nonlinear finite-element solver is initialized using the neural prediction. The results demonstrate that the proposed operator-learning framework can accurately capture coupled magneto-mechanical responses across a broad range of parametric problem settings.

\end{abstract}


\end{frontmatter}


\section{Introduction}
\label{sec:intro}

Magneto-active elastomers (MAEs) are a class of field-responsive smart composites consisting of magnetizable particles embedded in a soft polymeric matrix. When exposed to an external magnetic field, the interaction between magnetic particles and the surrounding elastomer can induce deformation and modify the effective mechanical response of the material. Owing to their tunable and remotely controllable behavior, MAEs have attracted considerable attention for applications in soft robotics, adaptive structures, vibration control, and smart actuation systems \cite{bira2020review, mondal2026magnetoactive, review1, aghamiri2025magnetoactive, aghamiri1, hossain2025recent,WANG2023116065}. From a continuum-mechanical perspective, their response is governed by a coupled magneto-mechanical boundary-value problem in which the deformation field and magnetic field are mutually dependent. The mechanical equilibrium equation must therefore be solved together with the governing magnetostatic equations, while the constitutive response couples the deformation gradient and magnetic field through the underlying free-energy function \cite{nadzharyan2022theoretical}. The numerical treatment becomes particularly demanding for heterogeneous MAEs, where large contrasts between the magnetic particles and the surrounding elastomer lead to strongly spatially varying material properties and localized field gradients \cite{klausler2025finite, klausler2026decoupled}.

The continuum-mechanical description of MAEs has evolved from early semi-empirical models to thermodynamically consistent finite-strain magneto-elastic formulations. A fundamental framework was established by Dorfmann and Ogden~\cite{dorfmann1}, in which the coupled response is described by a free-energy function depending on the deformation and magnetic variables. Furthermore, the mechanical stress and magnetic quantities follow consistently from derivatives of the same energy potential~\cite{dorfmann1}. Related variational formulations were subsequently developed for the treatment of finite-strain magneto-mechanical boundary-value problems \cite{kankanala2004finitely}. These theories have been extended to account for isotropic and anisotropic material responses, particle-chain-induced directionality, magnetic saturation, and microstructural effects \cite{bustamante2010transversely, danas2012experiments, saxena2015modelling}. Furthermore, finite-strain magneto-viscoelastic formulations have been introduced to describe the rate-dependent response of the polymeric matrix and magnetic relaxation under transient and cyclic loading \cite{saxena2013theory, saxena2014, DANAS_magnetovisco,RAMBAUSEK}. Together, these developments provide a comprehensive continuum framework for describing the strongly coupled, nonlinear, and microstructure-dependent response of MAEs. 

The resulting magneto-mechanical boundary-value problems are generally nonlinear and strongly coupled, particularly in the presence of finite deformations and heterogeneous material distributions. Since analytical solutions are rarely available for such problems, numerical techniques such as the finite-difference method (FDM), finite volume method (FVM), and finite element method (FEM) are commonly employed~\cite{Peiró2005, wu2022finite}. Although these approaches are robust and well-established, they require that the discretized governing equations be solved separately for each problem configuration, which can become computationally expensive for high-dimensional, nonlinear, and heterogeneous systems. This cost is particularly high in applications that require repeated simulations, thereby motivating the development of data-driven and physics-informed machine-learning approaches for efficiently approximating the underlying solution maps \cite{nooraiepour2026partial, brunton2024promising}.

In recent years, deep learning (DL), a branch of machine learning, has attracted increasing attention in computational mechanics and material modeling because of its ability to learn complex nonlinear relationships from data and provide rapid predictions after training \cite{herrmann2024deep}. Neural-network-based models have been successfully applied to tasks such as constitutive modeling \cite{dornheim2024neural, mouratidou2025neural}, multiscale analysis \cite{kim2024review, kalina2023fe}, microstructure–property relations, and full-field prediction of stress and deformation. In conventional supervised learning, however, the accuracy of these models depends strongly on the availability of sufficiently large and representative datasets, which are typically generated through experiments or high-fidelity numerical simulations. Moreover, purely data-driven models do not inherently satisfy the governing physical laws. These limitations have motivated the development of physics-informed neural networks, in which the governing equations and boundary conditions are incorporated directly into the training objective \cite{raissi2019physics}.

Physics-informed learning has further been extended to multiphysics and multiscale problems involving coupled systems of PDEs. In these settings, several physical fields are predicted simultaneously, and the corresponding governing equations are incorporated as coupled residual contributions in the training objective \cite{rezaei2022mixed, harandi2024mixed, Eshkofti02062024, Eshkofti04052026, wu2024pinn}. In this context, data-driven multiscale schemes have already been employed to replace repeated microscale magneto-mechanical RVE calculations by neural-network-based surrogate models of the effective material response \cite{KALINA2024116739, roth2026data, HARAZIN2026118726}. Gong et al.~\cite{10141630} applied PINNs to two-dimensional magnetostatic problems involving multiple magnetic media; however, their formulation was restricted to the electromagnetic field and did not account for mechanical deformation or magneto-mechanical coupling. More recently, Wang et al.~\cite{wang2024physics} developed a PINN framework for finite-deformation hyperelastic-magnetic problems in magnetorheological elastomers (MREs), where the magnetic and mechanical fields were coupled through the Maxwell stress tensor. Their formulation considered weak magneto-mechanical coupling, in which the magnetic field affects the deformation, while the deformation does not influence the magnetic field. In addition, the PINN was trained using a combination of physics-based residuals and a limited amount of labeled FEM data for specific boundary-value problems.

Despite their ability to incorporate governing equations directly into the learning process, conventional PINNs still face several challenges, including high training costs, strong sensitivity to network architecture and loss weighting, and difficulties in resolving sharp gradients or strongly coupled multiphysics responses. In addition, they are typically trained for a specific boundary-value problem and may require retraining when the geometry, material properties, loading, or boundary conditions change \cite{rezaei2022mixed, grossmann2024can, gultekin2025physics}.  More recently, physics-augmented neural constitutive models have been developed, where fundamental requirements such as objectivity, material symmetry, and thermodynamic consistency are incorporated into the neural representation \cite{ORTIGOSA2025117741, KLEIN2024116910}, while the resulting constitutive models are still embedded within a numerical solver to obtain the solution of each boundary-value problem.

These limitations motivate operator-learning approaches that aim to learn the solution map for an entire family of PDE problems rather than a single instance. Among the foundational architectures, DeepONet~\cite{lu2021learning} employs branch and trunk networks to approximate nonlinear operators, whereas the Fourier neural operator (FNO)~\cite{li2020fourier} parameterizes global integral operators efficiently in Fourier space. Alternative representations, including wavelet neural operators~\cite{tripura2023wavelet}, have also been proposed to improve the representation of localized spatial and frequency features. Following the initial developments of neural operators, increasing attention has been directed toward their use in computational mechanics. In this field, nonlinear constitutive behavior, heterogeneous material distributions, complex geometries, and multiscale interactions introduce substantially greater challenges~\cite{hildebrand2024comparison}. Operator-learning approaches have consequently been investigated for fracture mechanics~\cite{goswami2022physics}, nonlinear elasticity and elastoplasticity~\cite{koric2024deep,he2023novel}, heterogeneous microstructures~\cite{harandi2025spifol}, computational homogenization~\cite{zhang2024operator}, and geometry-dependent structural response~\cite{he2024geom}. For coupled multiphysics problems, this capability is especially valuable, as the learned operator can approximate the joint response of several interacting physical fields over a family of parameterized problem configurations \cite{ cai2021deepm,kushwaha2024advanced,rahman2024pretraining,park2026network,YAMAZAKI2026112132}. 

More recently, physics-informed neural operators have been developed to reduce or eliminate the dependence on labeled solution data by incorporating the governing physics directly into operator training. One important class of approaches enforces the governing equations through strong-form residuals evaluated by automatic differentiation, as demonstrated by physics-informed DeepONets~\cite{wang2021learning} and, more recently, latent-operator formulations~\cite{KARUMURI2026118599}. In parallel, variational and weak-form formulations have been increasingly explored to reduce the order of spatial differentiation required during training. For example, Eshaghi et al.~\cite{eshaghi2025variational} introduced the Variational Physics-Informed Neural Operator (VINO), in which parametric solution operators are trained by minimizing discretized variational functionals without labeled solution data. Finite operator learning (FOL)~\cite{REZAEI2026104506} further incorporated residuals derived from established numerical discretizations directly into the training objective, thereby avoiding automatic differentiation with respect to the spatial coordinates. This concept was subsequently extended through implicit finite operator learning (iFOL)~\cite{NAJIANASL2026108102} to obtain continuous parametric solution representations on arbitrary geometries. These developments reflect a broader evolution of physics-informed operator learning from strong-form, automatic-differentiation-based enforcement towards variational, weak-form, and numerical-method-guided formulations.

Despite these advances, physics-informed operator learning for strongly coupled finite-strain multiphysics problems remains comparatively less developed. While recent studies demonstrate the feasibility of neural operators for multi-field systems, the treatment of fully coupled finite-strain magneto-mechanical problems, in which the magnetic and mechanical fields mutually modify one another through the constitutive response, has received considerably less attention. Training physics-informed models with several simultaneously enforced physical objectives can be challenging because the corresponding loss terms may exhibit substantially different scales and gradient characteristics, resulting in poorly conditioned optimization landscapes and unbalanced training~\cite{wang2021gradient,krishnapriyan2021failure}. These
difficulties can become more pronounced in coupled multiphysics systems, where the individual solution fields are mutually dependent and the choice of network architecture has been shown to influence the representation of strongly coupled responses~\cite{park2026network}. For magneto-mechanical problems, additional numerical complexity arises from their coupled variational structure; in particular, mixed magneto-mechanical formulations can lead to saddle-point systems that are more difficult to solve numerically~\cite{vogel2013mixed,gebhart2024energy}. These
observations highlight the need for a coupled operator-learning formulation that preserves the physical interaction between the magnetic and mechanical fields while reducing interference between their respective optimization pathways.

To address these challenges, this study develops a data-free weak-form staggered neural operator (WSNO) framework for the finite-strain, fully coupled magneto-mechanical response of magneto-active elastomers. Building on the finite-operator-learning concept~\cite{REZAEI2026104506}, the governing magnetostatic and mechanical equations are enforced through finite-element weak-form residuals, thereby avoiding both labeled finite-element solution data and automatic differentiation with respect to the spatial coordinates. Separate field-specific neural operators are introduced for the magnetic potential and mechanical displacement fields. Rather than optimizing both operators simultaneously through a monolithic physics-informed objective, their trainable parameters are updated alternately using a staggered strategy. During each training stage, the current prediction of the complementary field is retained in the coupled constitutive relations and residual evaluation while being excluded from the corresponding backpropagation pathway. In this way, the physical magneto-mechanical coupling is retained while the optimization pathways of the two field-specific operators are separated.

The remainder of this paper is organized as follows. Section~2 introduces the finite-strain coupled magneto-mechanical formulation and the corresponding finite-element weak-form residuals used to construct the physics-informed loss. Section~3 presents the proposed WSNO framework, including the field-specific neural operators, the weak-form training objective, and the staggered optimization strategy. Section~4 evaluates the proposed framework through a series of numerical studies involving heterogeneous microstructures, variations in magnetic phase contrast, area-fraction-dependent geometries, strongly out-of-distribution material configurations combined with neural-initialized Newton correction, and a three-dimensional geometry-parametric problem. Finally, Section~5 summarizes the main findings and outlines directions for future study.

\section{Coupled partial differential equations and loss formulation}
\label{sec:coupled_pdes}

This section summarizes the coupled magneto-mechanical boundary-value problem employed throughout this study and introduces the corresponding finite-element-based physics-informed loss formulation. The governing equations are based on the finite-strain magneto-mechanical framework~\cite{kadapa2022unified, saber2023modeling}.  The coupled problem is formulated in terms of the mechanical displacement field $\boldsymbol{u}$ and the magnetic scalar potential $\phi$ as the primary unknowns. 

\subsection{Finite-strain magneto-mechanical problem}
\label{subsec:magnetomechanical_problem}

Under magnetostatic conditions and in the absence of free electric currents, Maxwell's equations in the reference configuration reduce to
\begin{equation}
\nabla_{\!X}\times\mathbb{H}
=
\boldsymbol{0},
\qquad
\nabla_{\!X}\cdot\mathbb{B}
=
0
\qquad
\text{in } \Omega_0,
\label{eq:magnetostatic_maxwell}
\end{equation}
where $\mathbb{H}$ and $\mathbb{B}$ denote the referential magnetic field and magnetic induction, respectively. Introducing the magnetic scalar potential $\phi$, the magnetic field can be measured as
\begin{equation}
\mathbb{H}
=
-\nabla_{\!X}\phi,
\label{eq:magnetic_scalar_potential}
\end{equation}
automatically satisfies the curl-free condition in Equation~\eqref{eq:magnetostatic_maxwell}. The remaining magnetic governing equation is therefore
\begin{equation}
\nabla_{\!X}\cdot
\mathbb{B}
=
0
\qquad
\text{in } \Omega_0.
\label{eq:magnetic_governing_equation}
\end{equation}

The magnetic boundary is decomposed into Dirichlet and Neumann parts, 
\begin{equation}
\Gamma_{\phi}
=
\Gamma_{\phi}^{D}
\cup
\Gamma_{\phi}^{N},
\qquad
\Gamma_{\phi}^{D}
\cap
\Gamma_{\phi}^{N}
=
\varnothing.
\end{equation}

The corresponding boundary conditions are prescribed as
\begin{equation}
\phi
=
\bar{\phi}
\qquad
\text{on } \Gamma_{\phi}^{D},
\label{eq:magnetic_dirichlet}
\end{equation}
and
\begin{equation}
\mathbb{B}\cdot\boldsymbol{N}
=
\mathbb{\bar{B}}_{N}
\qquad
\text{on } \Gamma_{\phi}^{N},
\label{eq:magnetic_neumann}
\end{equation}
where $\boldsymbol{N}$ denotes the outward unit normal in the reference configuration.

In the absence of inertia, the mechanical equilibrium equation is
\begin{equation}
\nabla_{\!X}\cdot
\boldsymbol{P}
\left(
\boldsymbol{F},
\mathbb{H}
\right)
+
\boldsymbol{b}_0
=
\boldsymbol{0}
\qquad
\text{in } \Omega_0,
\label{eq:mechanical_equilibrium}
\end{equation}
where $\boldsymbol{P}$ is the total first Piola-Kirchhoff stress and $\boldsymbol{b}_0$ denotes the mechanical body-force density in the reference configuration. Body forces are neglected in the numerical examples considered in this study.

The mechanical boundary is similarly decomposed as
\begin{equation}
\Gamma_u
=
\Gamma_u^D
\cup
\Gamma_u^N,
\end{equation}
with
\begin{equation}
\boldsymbol{u}
=
\bar{\boldsymbol{u}}
\qquad
\text{on } \Gamma_u^D,
\label{eq:mechanical_dirichlet}
\end{equation}
and
\begin{equation}
\boldsymbol{P}\boldsymbol{N}
=
\bar{\boldsymbol{t}}
\qquad
\text{on } \Gamma_u^N.
\label{eq:mechanical_neumann}
\end{equation}

The dependence of both $\boldsymbol{P}$ and $\mathbb{B}$ on the deformation and magnetic field gives rise to the coupled magneto-mechanical problem \cite{dorfmann2023nonlinear, ghosh2026constitutive}.

\subsection{Magneto-elastic constitutive response}
\label{subsec:magnetoelastic_constitutive}

The constitutive response is derived from a total magneto-elastic energy density function
\begin{equation}
\Psi
=
\Psi
\left(
\boldsymbol{F},
\mathbb{H};
\boldsymbol{X}
\right),
\label{eq:total_magnetoelastic_energy}
\end{equation}
where the explicit dependence on $\boldsymbol{X}$ allows for spatially heterogeneous material properties. The total energy is decomposed as
\begin{equation}
\Psi
=
\Psi^{\mathrm{mech}}
+
\Psi^{\mathrm{coup}}.
\label{eq:energy_decomposition}
\end{equation}
Here, $\Psi^{\mathrm{mech}}$ describes the elastic response of the elastomer, and $\Psi^{\mathrm{coup}}$ accounts for the coupling between deformation and the magnetic field. The magneto-mechanical response is described using both a linear magnetization contribution and a deformation-dependent coupling formulation. 

For the isotropic linear magnetic response considered in the present study, the coupled energy function is written as
\begin{equation}
\Psi_{\mathrm{isot}}^{\mathrm{coup}}(\mathbf{C},\mathbb{H})
=
-\frac{1}{2}\,\mu_0\mu_r
\mathbf{C}^{-1}
:
\left[\mathbb{H}\otimes\mathbb{H}\right],
\label{eq:linear_magnetic_response}
\end{equation}
where $\mu_0$ is the magnetic permeability of free space and $\mu_r$ is the local relative permeability~\cite{klausler2023meso}.

To characterize a deformation-dependent magnetic response of MAEs, the following formulation is used
\begin{equation}
I_6 = \mathbf{I} : \left[\mathbb{H} \otimes \mathbb{H}\right],
\qquad
I_7 = \mathbf{C} : \left[\mathbb{H} \otimes \mathbb{H}\right],
\qquad
\Psi_{\mathrm{isot}}^{\mathrm{coup}} = c_1 I_6 + c_2 I_7 .
\label{eq:nonlinear_magnetic_response}
\end{equation}
The invariant $I_6$ represents a purely magnetic response, and the invariant $I_7$ is used to consider the coupled magneto-mechanical contribution. Moreover, $c_1$ and $c_2$ are material parameters that govern the magnetic and coupled response, respectively~\cite{bustamante2010transversely}. 

The constitutive quantities follow from the energy potential as
\begin{equation}
\boldsymbol{P}
=
\frac{\partial\Psi}
{\partial\boldsymbol{F}},
\qquad
\mathbb{B}
=
-
\frac{\partial\Psi}
{\partial\mathbb{H}}.
\label{eq:constitutive_P_B}
\end{equation}

Accordingly, both the mechanical stress and magnetic induction depend on the coupled state variables $\boldsymbol{F}$ and $\mathbb{H}$. A change in the magnetic field therefore modifies the mechanical response, while deformation alters the magnetic constitutive response \cite{dorfmann2025constitutive}.

For the heterogeneous examples considered in this study, the local material response is parametrized in terms of the spatially varying Young's modulus $E(\boldsymbol{X})$ and relative magnetic permeability $\mu_r(\boldsymbol{X})$. Consequently, the total free-energy density depends explicitly on these heterogeneous material fields and can be written as
\begin{equation}
\Psi
=
\Psi
\left(
\boldsymbol{F},
\mathbb{H};
E(\boldsymbol{X}),
\mu_r(\boldsymbol{X})
\right).
\label{eq:heterogeneous_energy}
\end{equation}

The neural operators therefore learn mappings from heterogeneous material-property fields to the corresponding coupled magnetic and mechanical solution fields \cite{harandi2024mixed,REZAEI2026104506, rezaei2025finite}.

\subsection{Weak-form residual formulation}
\label{subsec:fe_residual_formulation}

To construct the physics-informed loss, the governing strong forms are first recast in their corresponding weak forms. Let $\delta\boldsymbol{u}$ and $\delta\phi$ denote admissible mechanical and magnetic test functions, respectively \cite{kanan2021finite}. In the reference configuration, the mechanical weak form is written as
\begin{equation}
G_u
=
\int_{\Omega_0}
\nabla_{\!X}\delta\boldsymbol{u}
:
\boldsymbol{P}
\,\mathrm{d}V
-
\int_{\Gamma_u^N}
\delta\boldsymbol{u}
\cdot
\bar{\boldsymbol{t}}
\,\mathrm{d}A
-
\int_{\Omega_0}
\delta\boldsymbol{u}
\cdot
\boldsymbol{b}_0
\,\mathrm{d}V
=
0,
\label{eq:mechanical_weak_form}
\end{equation}
where $\boldsymbol{P}$ denotes the total first Piola-Kirchhoff stress, $\bar{\boldsymbol{t}}$ is the prescribed traction on $\Gamma_u^N$, and $\boldsymbol{b}_0$ is the mechanical body-force density per unit reference volume.

The magnetic weak form is expressed as
\begin{equation}
G_{\phi}
=
\int_{\Omega_0}
\nabla_{\!X}\delta\phi
\cdot
\mathbb{B}
\,\mathrm{d}V
-
\int_{\Gamma_{\phi}^{N}}
\delta\phi
\,
\mathbb{\bar B_N}
\,\mathrm{d}A
=
0,
\label{eq:magnetic_weak_form}
\end{equation}
where $\mathbb{B}$ denotes the referential magnetic induction and $\mathbb{\bar B_N} = \mathbb{B}\cdot\boldsymbol{N}$ is the prescribed normal magnetic flux on $\Gamma_{\phi}^{N}$.

The displacement and magnetic scalar-potential fields are interpolated within each finite element as
\begin{equation}
\boldsymbol{u}^{e}
=
\boldsymbol{N}_{u}\boldsymbol{U}^{e},
\qquad
\phi^{e}
=
\boldsymbol{N}_{\phi}\boldsymbol{\Phi}^{e},
\label{eq:fe_interpolation}
\end{equation}
with the corresponding gradients
\begin{equation}
\nabla_{\!X}\boldsymbol{u}^{e}
=
\boldsymbol{B}_{u}\boldsymbol{U}^{e},
\qquad
\nabla_{\!X}\phi^{e}
=
\boldsymbol{B}_{\phi}\boldsymbol{\Phi}^{e},
\label{eq:fe_gradient_interpolation}
\end{equation}
where
\begin{equation}
\boldsymbol{B}_{u}
=
\nabla_{\!X}\boldsymbol{N}_{u},
\qquad
\boldsymbol{B}_{\phi}
=
\nabla_{\!X}\boldsymbol{N}_{\phi}.
\label{eq:shape_function_gradients}
\end{equation}
Here, $\boldsymbol{N}_{u}$ and $\boldsymbol{N}_{\phi}$ denote the shape-function matrices associated with the displacement and magnetic scalar-potential fields, respectively, while $\boldsymbol{B}_{u}$ and $\boldsymbol{B}_{\phi}$ are their corresponding gradient matrices with respect to the reference coordinates
$\boldsymbol{X}$.

The element-level mechanical residual in the reference configuration is written as
\begin{equation}
\boldsymbol{R}_u^e
=
\int_{\Omega_0^e}
\boldsymbol{B}_u^{T}
\boldsymbol{P}
\,\mathrm{d}V
-
\int_{\Gamma_u^{N,e}}
\boldsymbol{N}_u^{T}
\bar{\boldsymbol{t}}
\,\mathrm{d}A
-
\int_{\Omega_0^e}
\boldsymbol{N}_u^{T}
\boldsymbol{b}_0
\,\mathrm{d}V ,
\label{eq:mechanical_residual_integral}
\end{equation}
where $\Omega_0^e$ denotes the domain of element $e$ in the reference configuration, and $\Gamma_u^{N,e}$ denotes the part of its boundary subjected to prescribed mechanical tractions.

Similarly, the element-level magnetic residual is written as
\begin{equation}
\mathbf{R}_{\phi}^{e}
=
\int_{\Omega_0^{e}}
\mathbf{B}_{\phi}^{T}\,
\mathbb{B}
\,\mathrm{d}V
-
\int_{\Gamma_{\phi}^{N,e}}
\mathbf{N}_{\phi}^{T}\,
\mathbb{\bar{B}}_{N}
\,\mathrm{d}A .
\label{eq:magnetic_residual_integral}
\end{equation}

The residual contributions are subsequently assembled over all finite elements to obtain the global mechanical and magnetic residual vectors, $\boldsymbol{R}_u$ and $\boldsymbol{R}_{\phi}$, which are used to construct the physics-informed loss functions.

\section{Weak-form staggered neural operator framework}
\label{sec:WSNO}

The coupled magneto-mechanical formulation developed in Section~\ref{sec:coupled_pdes} provides governing equations used to train the neural operators without labeled solution data. In this study, a WSNO framework is used to learn the parametric mapping between continuous parameter spaces and the corresponding coupled magnetic and mechanical solution fields. In contrast to a conventional FE analysis, in which the nodal magnetic and mechanical degrees of freedom are solved independently for each problem setup, the WSNO learns an operator that approximates the solutions of a family of magneto-mechanical boundary-value problems after training. 

\begin{figure}[!ht]
    \centering
    \includegraphics[width=\textwidth]{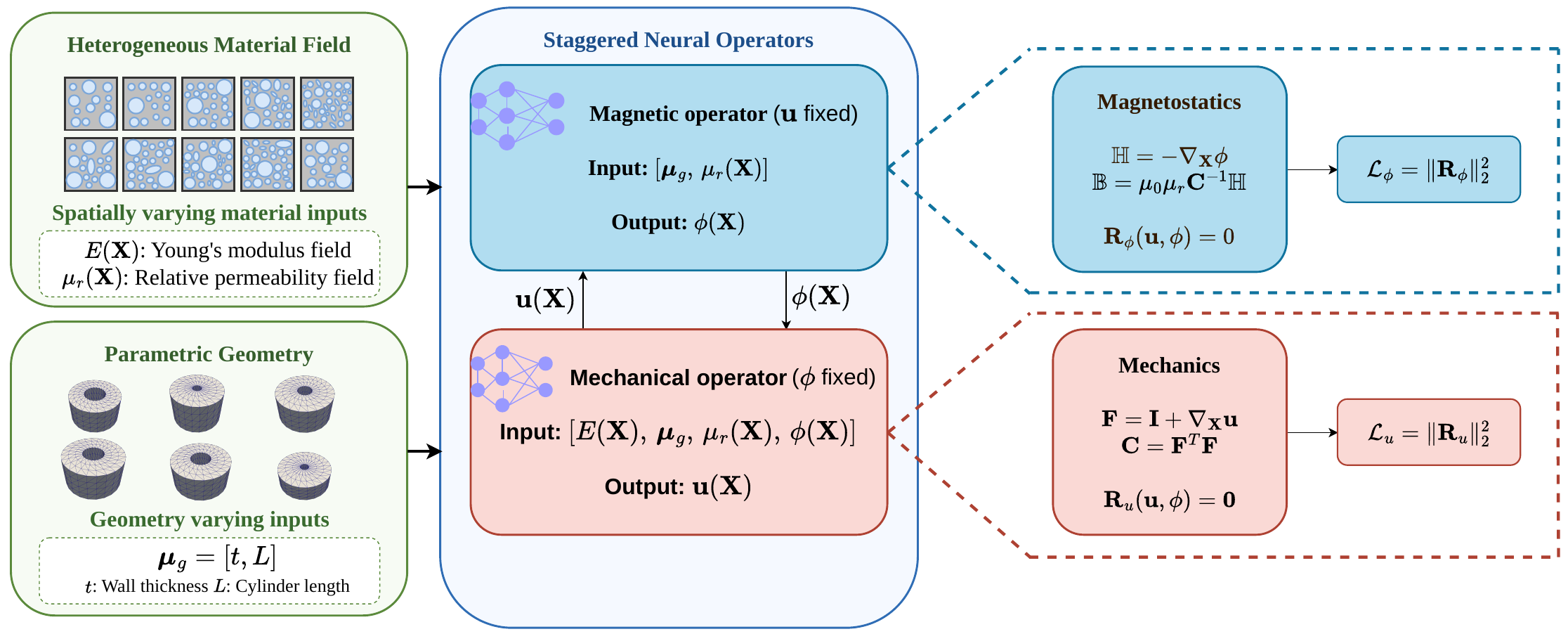}
    \caption{Schematic representation of the proposed WSNO framework for coupled magneto-mechanical problems. The magnetic and mechanical neural operators are trained alternately using the corresponding finite-element weak-form residuals while retaining the constitutive coupling between the magnetic potential and displacement fields.}
    \label{fig:wsno_architecture}
\end{figure}

The overall architecture of the proposed framework is shown in Figure~\ref{fig:wsno_architecture}. The WSNO is formulated as a parametric operator-learning framework, in which different problem realizations are represented through spatially varying input fields. In the heterogeneous magneto-active composite considered here, each realization corresponds to a different distribution of magnetic inclusions and is characterized by the associated Young's modulus \(E(\mathbf{X})\) and relative magnetic permeability \(\mu_r(\mathbf{X})\). The neural operators therefore learn the parametric mapping from these heterogeneous material distributions to the corresponding coupled magnetic and mechanical solution fields. Two field-specific neural operators are employed to represent the magnetic and mechanical solution maps. The magnetic operator predicts the scalar magnetic potential \(\phi(\mathbf{X})\), whereas the mechanical operator predicts the displacement field \(\mathbf{u}(\mathbf{X})\). The two operators remain physically coupled through the constitutive relations and the finite-element residuals introduced in Section~\ref{sec:coupled_pdes}.

The proposed WSNO framework is independent of a particular neural-operator architecture. In this framework, different operator-learning models, such as Fourier neural operators (FNO)~\cite{FNO1}, DeepONets~\cite{lu2021learning}, or implicit finite operator-learning architectures (iFOL)~\cite{NAJIANASL2026108102}, can be employed to represent the magnetic and mechanical solution maps. The main feature of the present study is the development of a general finite-element weak-form physics-informed framework. Within this framework, the magnetic and mechanical neural operators are optimized in a staggered manner, while the choice of the underlying neural-operator architecture remains flexible.

\subsection{Field-specific neural operators}
\label{subsec:field_specific_operators} 

In this framework, $\boldsymbol{\theta}_{\phi}$ and $\boldsymbol{\theta}_{u}$ are considered as trainable parameters of the magnetic and mechanical neural operators, respectively. The magnetic operator predicts the scalar magnetic-potential field according to \begin{equation} 
\phi_{\boldsymbol{\theta}_{\phi}}(\mathbf{X}) = \mathcal{G}_{\phi} \left[ \mu_r(\mathbf{X}); \boldsymbol{\theta}_{\phi} \right]. \label{eq:magnetic_operator} 
\end{equation} 

Although the displacement field is not directly supplied as an input to the magnetic operator in Equation~\eqref{eq:magnetic_operator}, the magnetic problem remains coupled to the deformation through the constitutive response. In particular, the magnetic induction ($\mathbb{B}$) is obtained from Equation~\eqref{eq:constitutive_P_B}, where the energy function $\Psi$ depends on both the deformation and magnetic field. Therefore, the magnetic residual generally depends on both coupled fields as 
\begin{equation} 
\mathbf{R}_{\phi} = \mathbf{R}_{\phi} \left( \mathbf{u}, \phi \right). \label{eq:magnetic_residual_coupling} 
\end{equation}

The mechanical operator receives the heterogeneous material-property fields together with the current magnetic-potential prediction and returns the displacement field, 
\begin{equation} 
\mathbf{u}_{\boldsymbol{\theta}_{u}}(\mathbf{X}) = \mathcal{G}_{u} \left[ E(\mathbf{X}), \mu_r(\mathbf{X}), \phi_{\boldsymbol{\theta}_{\phi}}(\mathbf{X}); \boldsymbol{\theta}_{u} \right]. \label{eq:mechanical_operator} 
\end{equation} 

From the predicted displacement field, the deformation gradient and the right Cauchy-Green tensor are evaluated as 
\begin{equation} 
\mathbf{F} = \mathbf{I} + \nabla_{\!X}\mathbf{u}_{\boldsymbol{\theta}_{u}}, \qquad \mathbf{C} = \mathbf{F}^{T}\mathbf{F}.
\label{eq:deformation_measures_wsno} 
\end{equation} 

The first Piola-Kirchhoff stress follows from the coupled magneto-elastic energy function according to Equation~\eqref{eq:constitutive_P_B}. Consequently, the mechanical equilibrium residual also depends on both solution fields,

\begin{equation} 
\mathbf{R}_{u} = \mathbf{R}_{u} \left( \mathbf{u}, \phi \right). \label{eq:mechanical_residual_coupling} 
\end{equation} 

Hence, the separate magnetic and mechanical operators do not represent two independent problems. Instead, information is exchanged between the two physical fields through the magneto-mechanical constitutive equations and the corresponding residual evaluations. At the global level, the WSNO approximates the parametric solution mapping 

\begin{equation} 
\mathcal{G}: \left\{ E(\mathbf{X}), \mu_r(\mathbf{X}) \right\} \longmapsto \left\{ \phi(\mathbf{X}), \mathbf{u}(\mathbf{X}) \right\}. \label{eq:wsno_global_mapping} 
\end{equation} 

This formulation is not restricted to a particular microstructure. More generally, the input fields represent the parameterized description of a family of magneto-mechanical problems, while the outputs correspond to their associated solution fields.

\subsection{Weak-form physics-informed loss} 
\label{subsec:wsno_loss}

The finite-element residual formulation introduced in Section~\ref{subsec:fe_residual_formulation} is directly used to construct the physics-informed training algorithm. The fields predicted by the neural operators are inserted into the constitutive equations and subsequently into the assembled magnetic and mechanical residual vectors, 
\begin{equation} 
\mathbf{R}_{\phi} \left( \mathbf{u}_{\boldsymbol{\theta}_{u}}, \phi_{\boldsymbol{\theta}_{\phi}} \right) = \mathbf{0}, \label{eq:wsno_mag_residual} 
\end{equation} 
and 
\begin{equation} 
\mathbf{R}_{u} \left( \mathbf{u}_{\boldsymbol{\theta}_{u}}, \phi_{\boldsymbol{\theta}_{\phi}} \right) = \mathbf{0}. \label{eq:wsno_mech_residual} 
\end{equation} 

The corresponding magnetic and mechanical loss functions are defined as 
\begin{equation} 
\mathcal{L}_{\phi} = \left\| \mathbf{R}_{\phi} \right\|_{2}^{2}, \qquad \mathcal{L}_{u} = \left\| \mathbf{R}_{u} \right\|_{2}^{2}. \label{eq:wsno_individual_losses} 
\end{equation} 

If both physical fields are optimized simultaneously, the corresponding monolithic loss can be written as 

\begin{equation} 
\mathcal{L}_{\mathrm{mono}} = w_{\phi}\mathcal{L}_{\phi} + w_{u}\mathcal{L}_{u}, 
\label{eq:wsno_monolithic_loss} 
\end{equation} 
where $w_{\phi}$ and $w_{u}$ denote weighting parameters associated with the magnetic and mechanical residuals, respectively. An important characteristic of the proposed formulation is that the spatial derivatives required for evaluating the governing equations are obtained from the finite-element interpolation introduced in Section~\ref{subsec:fe_residual_formulation}. Accordingly, the neural operators provide the predicted nodal fields, while the finite-element shape functions, shape-function gradients, quadrature rules, and element assembly are used to evaluate the corresponding weak-form residuals. Therefore, automatic differentiation with respect to the spatial coordinates is not required for constructing the physics-based loss. Furthermore, no reference finite-element solution fields are used during training. Finite-element solutions are employed only after training to evaluate the accuracy of the predicted fields. The resulting framework is therefore fully physics-driven.

\subsection{Staggered training of the coupled operators} \label{subsec:staggered_training} 
A direct simultaneous optimization of Equation~\eqref{eq:wsno_monolithic_loss} is found to exhibit poor convergence for the present magneto-mechanical problem. The difficulty comes from the coupled dependence of the residuals in Equations~\eqref{eq:wsno_mag_residual} and~\eqref{eq:wsno_mech_residual}, together with the different magnitudes and optimization characteristics of the magnetic and mechanical contributions. If the parameters of both neural operators are updated simultaneously, changes in one predicted field immediately modify both coupled residuals. This interaction can lead to competing optimization directions and unstable convergence. To improve the robustness of the training process, a staggered strategy is adopted. The magnetic and mechanical operators are optimized alternately, while the current prediction of the complementary physical field is kept fixed during each training stage. 

It is important to note that the stop-gradient operation preserves the current value of the fixed field while preventing gradient propagation through it during backpropagation. Therefore, the fixed field still contributes to the coupled constitutive equations and residual evaluation, although the parameters of its corresponding neural operator are not updated during that training stage. In this framework, the training procedure starts with a magnetic pretraining stage. At the beginning of training, the body is assumed to be undeformed, where $\mathbf{u}^{(0)} = \mathbf{0}$.
By using this undeformed configuration, the magnetic operator is trained to obtain an initial approximation of the scalar magnetic potential, 
\begin{equation} \phi^{(0)} = \mathcal{G}_{\phi} \left[ \mu_r; \boldsymbol{\theta}_{\phi}^{(0)} \right]. \label{eq:magnetic_pretraining_wsno} 
\end{equation} 

Thereafter, in the staggered scheme, the resulting magnetic solution is employed to initialize the mechanical training stage.

\subsubsection{Mechanical training stage} 
During each mechanical training stage, the current magnetic-potential prediction is kept fixed using the stop-gradient operation as 
\begin{equation} 
\widehat{\phi}^{(k)} = \operatorname{sg} \left( \phi_{\boldsymbol{\theta}_{\phi}}^{(k)} \right). \label{eq:stop_gradient_phi}
\end{equation} 
Here, $\operatorname{sg}(\cdot)$ denotes the stop-gradient operator. It leaves the numerical value of its argument unchanged but prevents gradients from being propagated through that quantity during backpropagation. 

The mechanical operator therefore evaluates 
\begin{equation} 
\mathbf{u}_{\boldsymbol{\theta}_{u}} = \mathcal{G}_{u} \left[ E, \mu_r, \widehat{\phi}^{(k)}; \boldsymbol{\theta}_{u} \right]. \label{eq:mechanical_stage_prediction} 
\end{equation} 

During this stage, only the mechanical loss is minimized with respect to the mechanical network parameters, 
\begin{equation} 
\boldsymbol{\theta}_{u}^{(k+1)} = \underset{\boldsymbol{\theta}_{u}}{\operatorname{arg\,min}} \; \mathcal{L}_{u} \left( \mathbf{u}_{\boldsymbol{\theta}_{u}}, \widehat{\phi}^{(k)} \right). \label{eq:mechanical_stage_optimization} 
\end{equation} 

Although $\widehat{\phi}^{(k)}$ is excluded from backpropagation, it is still used to evaluate the magnetic field and the coupled stress response. Therefore, the mechanical problem remains physically coupled to the magnetic solution.

\subsubsection{Magnetic training stage} 
After completing the mechanical stage, the updated displacement prediction is kept fixed, 
\begin{equation} 
\widehat{\mathbf{u}}^{(k+1)} = \operatorname{sg} \left( \mathbf{u}_{\boldsymbol{\theta}_{u}}^{(k+1)} \right). \label{eq:stop_gradient_u} 
\end{equation} 

Then, the fixed displacement field is used to evaluate the deformation-dependent magnetic constitutive response. During this stage, only the magnetic loss is minimized, 
\begin{equation} \boldsymbol{\theta}_{\phi}^{(k+1)} = \underset{\boldsymbol{\theta}_{\phi}}{\operatorname{arg\,min}} \; \mathcal{L}_{\phi} \left( \phi_{\boldsymbol{\theta}_{\phi}}, \widehat{\mathbf{u}}^{(k+1)} \right). \label{eq:magnetic_stage_optimization} 
\end{equation} 

It is important to note that the displacement field $\widehat{\mathbf{u}}^{(k+1)}$ participates in the evaluation of the deformation gradient, constitutive response, magnetic induction, and magnetic residual. The stop-gradient operation therefore affects only the optimization pathway and does not remove the underlying magneto-mechanical coupling.

\subsubsection{Staggered training sequence} 
The complete staggered procedure can be summarized as 
\begin{equation} \mathbf{u}^{(0)} = \mathbf{0} \quad \longrightarrow \quad \phi^{(0)} \quad \longrightarrow \quad \mathbf{u}^{(1)} \quad \longrightarrow \quad \phi^{(1)} \quad \longrightarrow \quad \mathbf{u}^{(2)} \quad \longrightarrow \quad \phi^{(2)} \quad \longrightarrow \quad \cdots . 
\label{eq:wsno_staggered_sequence} 
\end{equation} 
At each substage, one neural operator is optimized while the current prediction of the other field is retained as a fixed coupled state. This procedure reduces simultaneous gradient interactions between magnetic and mechanical losses and is found to provide a substantially more stable training process than direct monolithic optimization. The term \emph{weak-form staggered neural operator} therefore indicates the three main components of the proposed framework. First, the governing magneto-mechanical equations are enforced using finite-element weak-form residuals. Second, the magnetic and mechanical fields are represented by neural operators that learn mappings over a parameterized family of boundary-value problems. Third, the two field-specific operators are trained alternately through a staggered optimization strategy while retaining the physical coupling between the magnetic and mechanical fields. Figure~\ref{fig:wsno_architecture} provides a schematic overview of this procedure. The heterogeneous material-property and geometry fields define the parametric input space, the magnetic and mechanical operators predict the corresponding solution fields, and the finite-element magnetostatic and mechanical residuals provide the physics-informed training objectives. The exchange of the predicted magnetic potential and displacement fields between the two stages preserves the coupled magneto-mechanical response throughout the staggered training procedure. 
The staggered training procedure used in the WSNO framework is summarized in Algorithm~\ref{alg:wsno_staggered}. The magnetic and mechanical operators are optimized alternately while retaining their coupled interaction through the weak-form residuals.

\section{Results}
\label{sec:results}

To evaluate the proposed WSNO framework, several magneto-mechanical benchmark problems are considered. The studies assess the prediction accuracy, generalization capability, and robustness of the learned operators under variations in microstructure, material distribution, and geometry. All geometric, mechanical, and magnetic quantities considered in the
following numerical examples are expressed in normalized form.
The normalization is defined in Appendix~\ref{app:normalization} in terms
of arbitrary characteristic length and mechanical stiffness scales,
$L_0$ and $E_0$, respectively. For notational simplicity, the overbars denoting normalized quantities are omitted throughout this section. The neural-operator predictions are evaluated against independently computed nonlinear finite-element solutions, which are used exclusively as reference data for validation and are not employed during training. The following studies examine random heterogeneous inclusion microstructures, area-fraction-dependent responses, out-of-distribution material configurations combined with neural-initialized Newton correction, and geometry-parametric three-dimensional magneto-mechanical problems.
In the numerical studies presented in this section, FNO-based operators are employed for the heterogeneous two-dimensional examples and the three-dimensional geometry-parametric problem.
A summary of the neural-operator hyperparameters, training settings, and problem-specific numerical parameters employed in the different benchmark studies is provided in~\ref{app:hyperparameters}.

\begin{algorithm}[htbp]
\caption{Condensed WSNO training procedure.}
\label{alg:wsno_staggered}
\small
\begin{algorithmic}[1]

\Require Training set $\mathcal{B}$, staggered cycles $N_{\mathrm{cyc}}$,
epochs $N_{\phi}^{\mathrm{pre}},N_u,N_{\phi},N_{\mathrm{ft}}$
\Ensure Trained parameters $\theta_{\phi},\theta_u$

\State Initialize $\theta_{\phi},\theta_u$ and set $\mathbf{u}^{(0)}=\mathbf{0}$

\Statex \textbf{Magnetic pretraining}
\For{$n=1,\ldots,N_{\phi}^{\mathrm{pre}}$}
    \State $\phi^{(0)} \gets \mathcal{G}_{\phi}(\mu_r;\theta_{\phi})$
    \State $\theta_{\phi}
    \gets
    \theta_{\phi}
    -\eta_{\phi}\nabla_{\theta_{\phi}}
    \left\|\mathbf{R}_{\phi}(\mathbf{u}^{(0)},\phi^{(0)})\right\|_2^2$
\EndFor

\Statex \textbf{Staggered training}
\For{$k=0,\ldots,N_{\mathrm{cyc}}-1$}

    \State $\widehat{\phi}^{(k)}
    \gets \operatorname{sg}\!\left(\phi^{(k)}\right)$
    \For{$n=1,\ldots,N_u$}
        \State $\mathbf{u}^{(k+1)}
        \gets \mathcal{G}_u(E,\mu_r,\widehat{\phi}^{(k)};\theta_u)$
        \State $\theta_u
        \gets
        \theta_u
        -\eta_u\nabla_{\theta_u}
        \left\|\mathbf{R}_u
        (\mathbf{u}^{(k+1)},\widehat{\phi}^{(k)})\right\|_2^2$
    \EndFor

    \State $\widehat{\mathbf{u}}^{(k+1)}
    \gets \operatorname{sg}\!\left(\mathbf{u}^{(k+1)}\right)$
    \For{$n=1,\ldots,N_{\phi}$}
        \State $\phi^{(k+1)}
        \gets \mathcal{G}_{\phi}(\mu_r;\theta_{\phi})$
        \State $\theta_{\phi}
        \gets
        \theta_{\phi}
        -\eta_{\phi}\nabla_{\theta_{\phi}}
        \left\|\mathbf{R}_{\phi}
        (\widehat{\mathbf{u}}^{(k+1)},\phi^{(k+1)})\right\|_2^2$
    \EndFor

\EndFor

\Statex \textbf{Optional coupled fine-tuning}
\For{$n=1,\ldots,N_{\mathrm{ft}}$}
    \State $\phi\gets\mathcal{G}_{\phi}(\mu_r;\theta_{\phi})$,
           $\mathbf{u}\gets\mathcal{G}_u(E,\mu_r,\phi;\theta_u)$
    \State $\mathcal{L}_{\mathrm{mono}}
    \gets
    w_{\phi}\|\mathbf{R}_{\phi}(\mathbf{u},\phi)\|_2^2
    +
    w_u\|\mathbf{R}_u(\mathbf{u},\phi)\|_2^2$
    \State Update $\theta_{\phi}$ and $\theta_u$ using
    $\nabla\mathcal{L}_{\mathrm{mono}}$
\EndFor

\State \Return $\theta_{\phi},\theta_u$

\end{algorithmic}
\end{algorithm}

\subsection{Coupled magneto-mechanical prediction for random inclusion microstructures}
\label{subsec:magneto_mechanical_inclusion}

To evaluate the capability of the proposed WSNO framework for coupled magneto-mechanical problems, a two-dimensional heterogeneous square domain containing randomly distributed circular inclusions is considered. The computational domain is defined as
\begin{equation}
\Omega = [0,L]\times[0,L],
\label{eq:mm_inclusion_domain}
\end{equation}
where \(L=1~[-]\). The domain is discretized using a \(42\times42\) structured quadrilateral mesh.

Each microstructure contains \(N_{\mathrm{inc}}\) circular inclusions. The geometry of the inclusions is described by the control vector
\begin{equation}
\boldsymbol{\mu}
=
\left[
c_{x,1},c_{y,1},r_1,
c_{x,2},c_{y,2},r_2,
\ldots,
c_{x,N_{\mathrm{inc}}},c_{y,N_{\mathrm{inc}}},r_{N_{\mathrm{inc}}}
\right]^T,
\label{eq:inclusion_control_vector}
\end{equation}
where \(c_{x,i}\), \(c_{y,i}\), and \(r_i\) denote the center coordinates and radius of the \(i\)-th inclusion, respectively. In this study, \(N_{\mathrm{inc}}=15\), and the inclusion radii are sampled within the range
\begin{equation}
r_i \in [0.05,0.10].
\label{eq:inclusion_radius_range}
\end{equation}

To obtain a differentiable representation of the material layout, each inclusion is described by a smooth indicator function. For the \(i\)-th inclusion, the indicator function is defined as
\begin{equation}
w_i(\mathbf X)
=
\mathrm{sigmoid}
\left[
\beta
\left(
r_i-\left\|\mathbf X-\mathbf c_i\right\|
\right)
\right],
\label{eq:single_inclusion_indicator}
\end{equation}
where \(\mathbf c_i=[c_{x,i},c_{y,i}]^T\) is the inclusion center and \(\beta\) controls the sharpness of the particle-matrix transition. The total inclusion indicator is obtained as
\begin{equation}
w(\mathbf X)
=
\mathrm{clip}
\left(
\sum_{i=1}^{N_{\mathrm{inc}}} w_i(\mathbf X),0,1
\right).
\label{eq:total_inclusion_indicator}
\end{equation}
The clipping operation prevents the indicator from exceeding unity in possible overlap regions.

The spatial distributions of Young's modulus and relative permeability are interpolated between the matrix and inclusion phases according to
\begin{equation}
E(\mathbf X)
=
E_m
+
w(\mathbf X)
\left(
E_i-E_m
\right),
\label{eq:young_modulus_interpolation_inclusion}
\end{equation}
and
\begin{equation}
\mu_r(\mathbf X)
=
\mu_{r,m}
+
w(\mathbf X)
\left(
\mu_{r,i}-\mu_{r,m}
\right).
\label{eq:relative_permeability_interpolation_inclusion}
\end{equation}
Here, the matrix properties are \(E_m=1~[-]\) and \(\mu_{r,m}=1\), while the inclusion properties are \(E_i=10~[-]\) and \(\mu_{r,i}=5\).

The magnetic scalar potential is prescribed on the left and right boundaries as
\begin{equation}
\phi = 0.15~[-]
\quad \mathrm{on} \quad
\Gamma_{\mathrm{left}},
\label{eq:mm_phi_left_bc}
\end{equation}
and
\begin{equation}
\phi = 0~[-]
\quad \mathrm{on} \quad
\Gamma_{\mathrm{right}}.
\label{eq:mm_phi_right_bc}
\end{equation}

\begin{figure}[!htbp]
\centering
\includegraphics[width=0.8\textwidth]{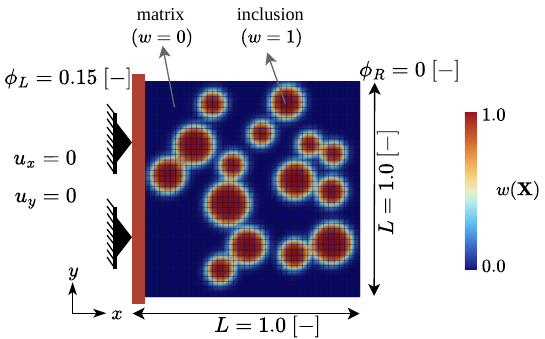}
\caption{Computational domain, boundary conditions, and representative random-inclusion microstructure. The magnetic scalar potential is prescribed as $\phi_L = 0.15~[-]$ and $\phi_R = 0~[-]$, while the displacement components $u_x$ and $u_y$ are fixed on the left boundary. The color map represents the combined smooth phase-indicator field $w(\mathbf{X})$.}
\label{fig:BC_micorstructure}
\end{figure}

\begin{figure}[!ht]
\centering
\includegraphics[angle=90,width=\textwidth]{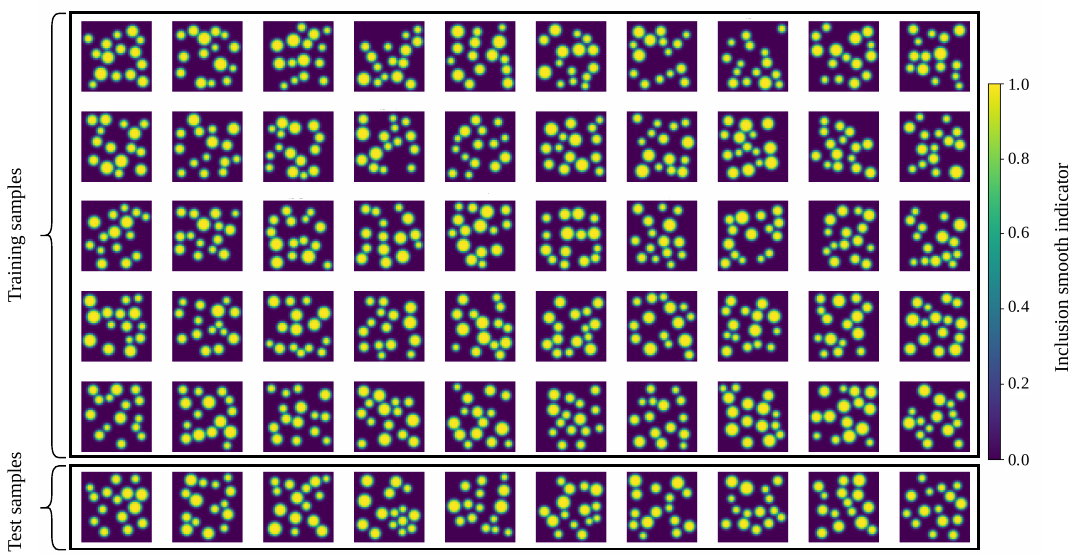}
\caption{Training and test microstructures for the coupled magneto-mechanical inclusion example. The first 50 samples are used for training, and 10 unseen samples are used for testing.}
\label{fig:training_test_microstructures}
\end{figure}

The prescribed magnetic scalar-potential and mechanical boundary conditions are illustrated in Figure~\ref{fig:BC_micorstructure}. The Figure~\ref{fig:training_test_microstructures} also shows a representative random distribution of circular inclusions in the elastomeric matrix together with the corresponding smooth phase-indicator field defined in Equations~\eqref{eq:single_inclusion_indicator} and~\eqref{eq:total_inclusion_indicator}.

The WSNO model learns the parametric solution operator
\begin{equation}
\mathcal{G}:
\left\{
E(\mathbf X),\mu_r(\mathbf X)
\right\}
\mapsto
\left\{
\phi(\mathbf X),u_x(\mathbf X),u_y(\mathbf X)
\right\}.
\label{eq:mm_pi_fno_operator}
\end{equation}

The magnetic field is recovered from the predicted scalar potential as
\begin{equation}
\mathbb{H}
=
-\nabla \phi.
\label{eq:mm_magnetic_field}
\end{equation}

A key difficulty in this example is the saddle-point structure of the coupled magneto-mechanical problem. As described in Section~\ref{subsec:staggered_training}, a direct monolithic minimization of the fully coupled physics-informed loss may become unstable or difficult to optimize. This issue is particularly relevant because the magnetic and mechanical residuals can have different magnitudes and sensitivities during training. Therefore, instead of training the fully coupled system in a single monolithic step, a staggered physics-informed training strategy is adopted. In this example, the staggered training procedure consists of three main stages (\ref{app:hyperparameters}). 

A total of 50 randomly generated microstructures is used for training, while 10 unseen microstructures are used for testing. The training and test microstructures are shown in Figure~\ref{fig:training_test_microstructures}.

The prediction accuracy is evaluated separately for the magnetic and mechanical fields. The relative \(L_2\) error of the magnetic scalar potential is defined as
\begin{equation}
e_{\phi}^{L_2}
=
\frac{
\left\|
\phi_{\mathrm{WSNO}}
-
\phi_{\mathrm{FE}}
\right\|_2
}{
\left\|
\phi_{\mathrm{FE}}
\right\|_2
}
\times 100.
\label{eq:mm_rel_l2_phi}
\end{equation}

The relative \(L_2\) error of the magnetic field is computed as
\begin{equation}
e_{\mathbb H}^{L_2}
=
\frac{
\left\|
\mathbb{H_{\mathrm{WSNO}}}
-
\mathbb{H_{\mathrm{FE}}}
\right\|_2
}{
\left\|
\mathbb{H_{\mathrm{FE}}}
\right\|_2
}
\times 100.
\label{eq:mm_rel_l2_h}
\end{equation}

Component-wise displacement errors are also evaluated as
\begin{equation}
e_{u_x}^{L_2}
=
\frac{
\left\|
u_{x,\mathrm{WSNO}}
-
u_{x,\mathrm{FE}}
\right\|_2
}{
\left\|
u_{x,\mathrm{FE}}
\right\|_2
}
\times 100.
\label{eq:mm_rel_l2_ux}
\end{equation}

The finite element solution is used only as a reference for evaluation and error measurement. The WSNO model predicts the magnetic scalar potential and displacement fields directly, while the magnetic field is obtained from the gradient of the predicted scalar potential. Therefore, the magnetic-field error is expected to be larger than the potential error because gradient-based quantities are more sensitive to local prediction errors.

\begin{figure}[!ht]
    \centering
    \includegraphics[width=\textwidth]{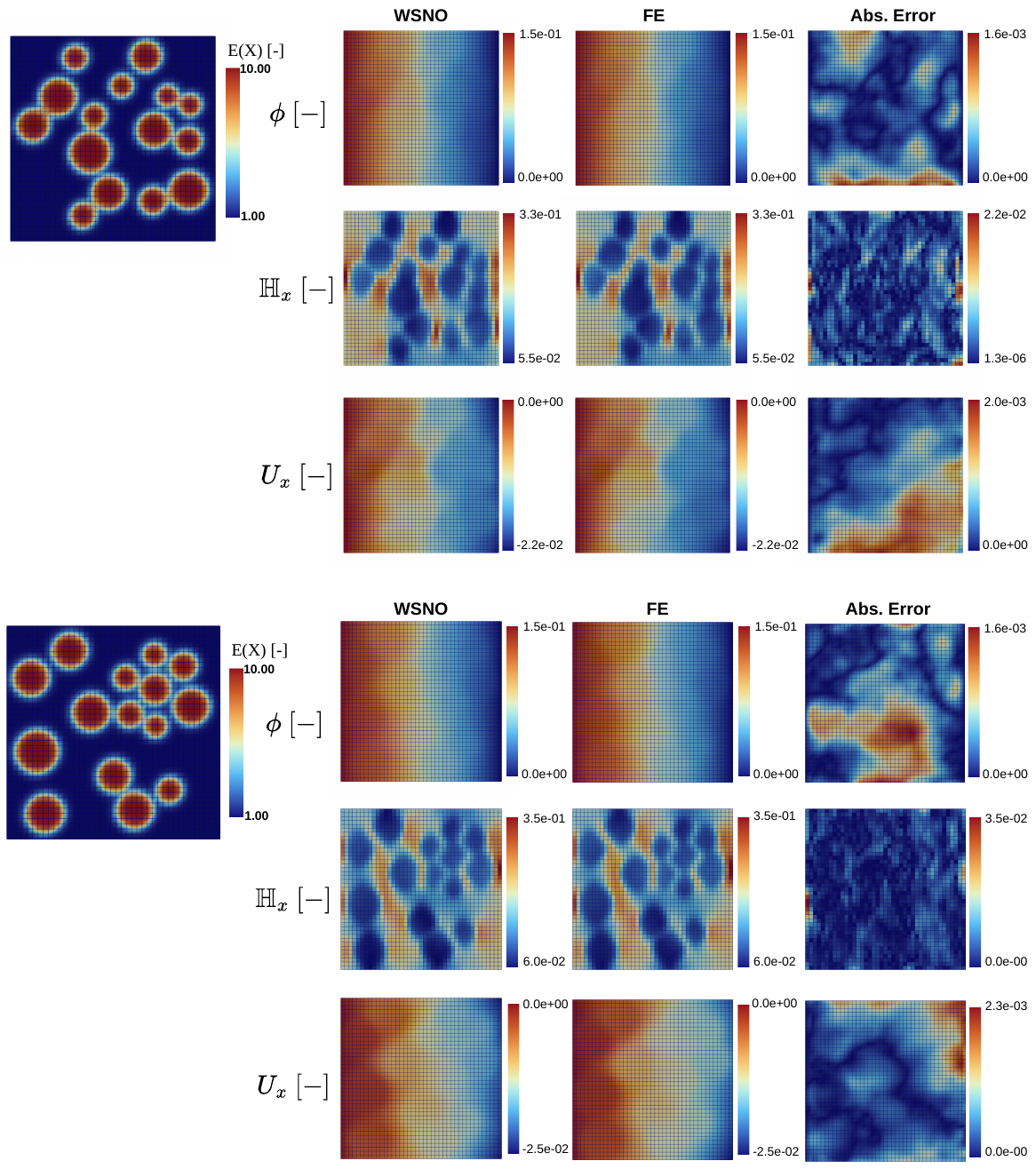}
    \caption{
    Representative WSNO prediction for two unseen random-inclusion
    microstructures.}
    \label{fig:random_inclusion_representative}
\end{figure}

Figure~\ref{fig:random_inclusion_representative} presents a representative prediction for two unseen random-inclusion microstructures. The prescribed heterogeneous elasticity field is shown. This figure compares the full-field WSNO predictions with the finite-element reference solutions for the magnetic scalar potential $\phi$, the magnetic-field component $\mathbb{H}_x$, and the displacement component $u_x$. The WSNO accurately reproduces the dominant spatial features of all investigated fields. The agreement is particularly strong for the scalar potential and displacement fields, while more localized discrepancies are observed for the magnetic-field components.
The local errors in $\mathbb{H}_x$ are consistent with the gradient-based recovery of the magnetic field based on Equation~\eqref{eq:magnetic_scalar_potential}, such that small local deviations in the predicted scalar potential can be amplified by spatial differentiation. Nevertheless, the absolute-error maps indicate that the discrepancies remain spatially localized, whereas the overall coupled magneto-mechanical field distributions are well captured by the trained WSNO model.

\begin{figure}[!htbp]
\centering
\includegraphics[width=0.5\textwidth]{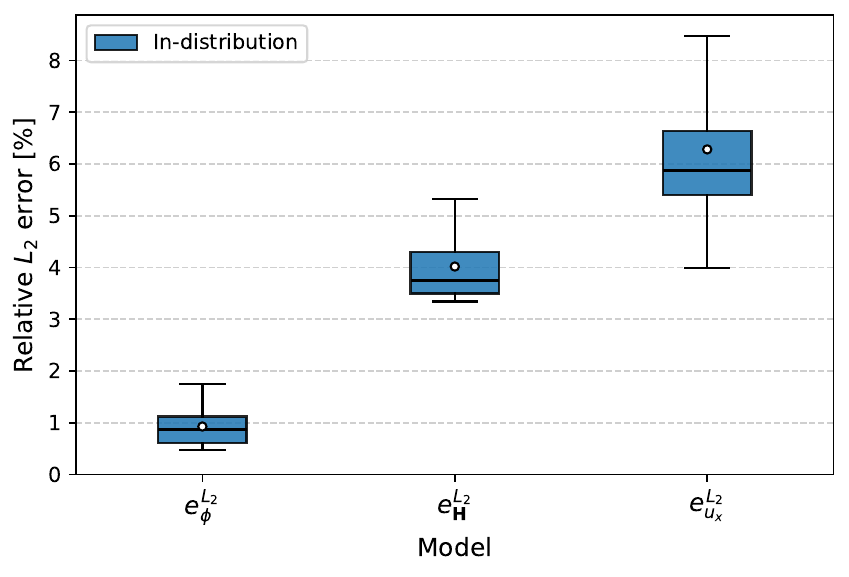}
\caption{Comparison of the relative \(L_2\) errors over 10 in-distribution test samples for the coupled magneto-mechanical inclusion problem. The errors are reported for the magnetic scalar potential, magnetic field, and horizontal displacement component \(u_x\).}
\label{fig:mm_l2_error_boxplot_phi_h_ux}
\end{figure}

Figure~\ref{fig:mm_l2_error_boxplot_phi_h_ux} shows the distribution of the relative \(L_2\) errors over 10 unseen in-distribution test samples. The errors are reported for the magnetic scalar potential \(\phi\), the magnetic field \(\mathbb{H}\), and the horizontal displacement component \(u_x\). The scalar potential is predicted with high accuracy, while the magnetic field exhibits a larger error because it is recovered from the spatial gradient of \(\phi\). The \(u_x\) error represents the mechanical prediction accuracy in the dominant displacement direction.

The results demonstrate that the proposed staggered WSNO framework can capture the coupled magneto-mechanical response of heterogeneous inclusion microstructures. The magnetic scalar potential is predicted with high accuracy, while the displacement field is more sensitive to local microstructural variations and inclusion-induced deformation patterns. Overall, this example shows that the staggered WSNO strategy provides a stable and efficient surrogate modeling approach for coupled magneto-mechanical problems with saddle-point structure.

\FloatBarrier
\subsection{Generalization across magnetic phase contrast}
\label{subsubsec:phase_contrast}

To further assess the parametric generalization capability of the WSNO framework, the random-inclusion study is extended by allowing the magnetic phase contrast between the inclusion and matrix phases to vary. In the Section~\ref{subsec:magneto_mechanical_inclusion}, the relative magnetic permeability of each phase is kept fixed, such that the parametric variation arose solely from changes in the microstructural configuration, including the positions and sizes of the inclusions. In the present study, both the microstructural configuration and the inclusion relative magnetic permeability are treated as parametric variables. The geometry generation procedure, computational domain, finite-element discretization, boundary conditions, and mechanical material properties are kept identical to those introduced in the random-inclusion study.

The matrix relative permeability is fixed at $\mu_{r,m}=1$, whereas the inclusion relative permeability is varied. The magnetic phase contrast is therefore characterized by
\begin{equation}
\chi_{\mu}
=
\frac{\mu_{r,i}}{\mu_{r,m}}
=
\mu_{r,i}.
\end{equation}

The WSNO is trained using three inclusion relative-permeability values,
\begin{equation}
\mu_{r,i}^{\mathrm{train}}
\in
\{2,5,10\}.
\end{equation}

The trained model is subsequently evaluated on ten unseen random-inclusion microstructures
at
\begin{equation}
\mu_{r,i}^{\mathrm{eval}}
\in
\{4,5,8\}.
\end{equation}

The case $\mu_{r,i}=5$ corresponds to a material contrast encountered during training and therefore evaluates generalization to unseen microstructures at a known permeability value. In contrast, $\mu_{r,i}=4$ and $8$ are unseen intermediate values within the training interval and are used to assess interpolation with respect to the magnetic phase contrast. Moreover, to investigate the influence of the training-set size, the study is conducted with two different numbers of training samples, namely 50 and 100 microstructures. It is worth noting that the same 10 test samples are used in all investigations to ensure a consistent comparison. 

\begin{figure}[htbp]
    \centering
    \includegraphics[width=\textwidth]{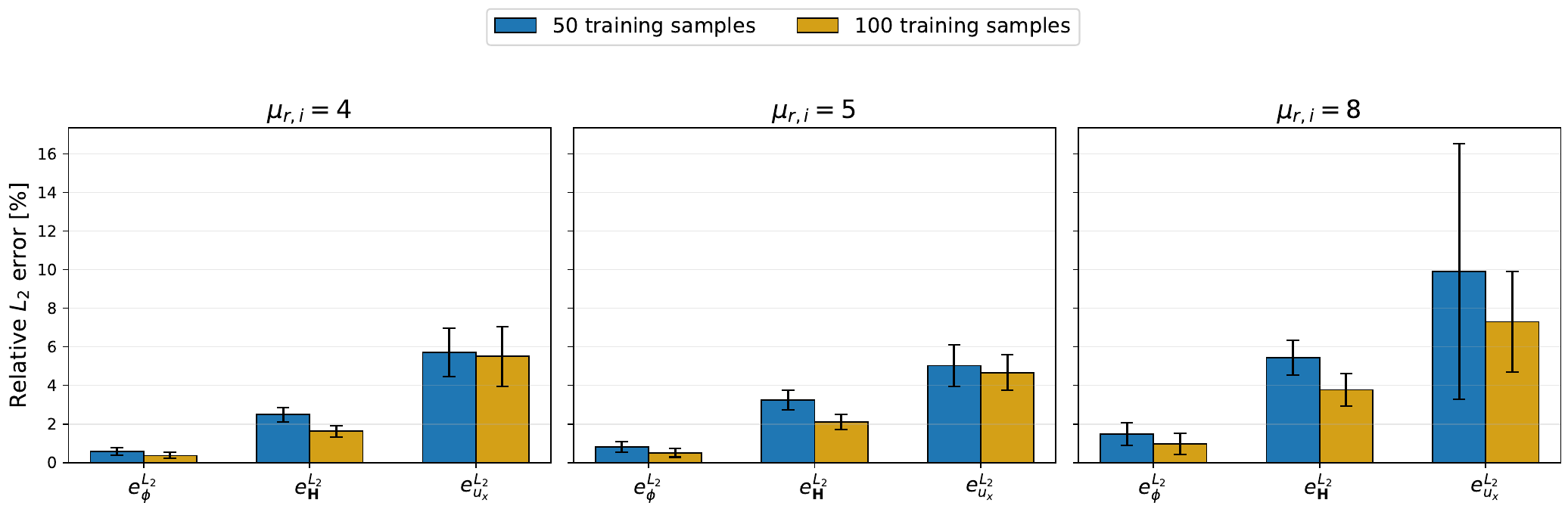}
    \caption{Comparison of the relative $L_2$ errors obtained using 50 and 100 training microstructures for different inclusion relative permeabilities $\mu_{r,i}=4$, $5$, and $8$. The errors are reported for the magnetic scalar potential $\phi$, magnetic field $\mathbb{H}$, and horizontal displacement component $u_x$. The bars represent the mean relative $L_2$ error over the 10 test microstructures, while the error bars indicate one standard
    deviation.}
    \label{fig:phase_contrast_training_size}
\end{figure}

Figure~\ref{fig:phase_contrast_training_size} shows the prediction accuracy of the implemented WSNO framework trained using 50 and 100 microstructures. The improvement in prediction accuracy with an increasing number of training samples is evident. The reduction in the scalar-potential error remains comparatively modest because $\phi$ is already predicted with high accuracy using 50 training samples. More noticeable improvements are observed for both the magnetic field $\mathbb{H}$ and the dominant displacement component $u_x$ across all considered relative-permeability contrasts. Overall, both the magnetic and mechanical responses exhibit acceptable relative errors with respect to the finite-element reference solutions, demonstrating the capability of the proposed framework to predict coupled responses across varying microstructures and magnetic phase contrasts.

Overall, these results indicate that the learned WSNO is not restricted to a single magnetic phase contrast. In addition to generalizing to unseen inclusion arrangements, the operator retains predictive capability at intermediate permeability values that are not explicitly included during training.

\FloatBarrier
\subsection{Area-fraction-dependent magneto-mechanical response of a single inclusion}
\label{subsec:area_fraction_single_inclusion}

To investigate the capability of the proposed framework for geometry parametric coupled problems, a square domain containing a single circular magneto-mechanical inclusion is considered. In contrast to the previous random-inclusion example, the present case investigates the effect of the inclusion size by using the inclusion area fraction as the only control parameter. 

The computational domain is defined as
\begin{equation}
\Omega=[0,L]\times[0,L],
\label{eq:single_inclusion_domain}
\end{equation}

where \(L=1~[-]\). A circular inclusion is placed at the center of the domain,

\begin{equation}
\mathbf X_c=
\begin{bmatrix}
L/2\\
L/2
\end{bmatrix}.
\label{eq:single_inclusion_center}
\end{equation}

The scalar control parameter is the inclusion area fraction \(f_A\). For a prescribed area fraction \(f_A\), the inclusion radius is computed as

\begin{equation}
r
=
\sqrt{
\frac{f_A L^2}{\pi}
}.
\label{eq:single_inclusion_radius}
\end{equation}

Thus, the model learns a one-dimensional geometry-to-solution operator.

To obtain a differentiable description of the particle-matrix interface, a smooth inclusion indicator based on Equation~\eqref{eq:single_inclusion_indicator} is used. Moreover, the spatially varying Young's modulus and the relative permeability are defined according to Equations~\eqref{eq:young_modulus_interpolation_inclusion} and~\eqref{eq:relative_permeability_interpolation_inclusion}. 

In this study, the matrix properties are \(E_m=1~[-]\) and \(\mu_{r,m}=1\), whereas the inclusion properties are \(E_i=10~[-]\) and \(\mu_{r,i}=10\).

The magnetic scalar potential is prescribed on the left and right boundaries as
\begin{equation}
\phi=0.15~[-]
\quad \text{on} \quad
\Gamma_{\mathrm{left}},
\qquad
\phi=0~[-]
\quad \text{on} \quad
\Gamma_{\mathrm{right}}.
\label{eq:single_inclusion_phi_bc}
\end{equation}

For the mechanical problem, the displacement components are fixed on the left boundary,
\begin{equation}
u_x=0,
\qquad
u_y=0
\quad \text{on} \quad
\Gamma_{\mathrm{left}}.
\label{eq:single_inclusion_u_bc}
\end{equation}

The WSNO learns the parametric solution operator
\begin{equation}
\mathcal{G}:
f_A
\mapsto
\left\{
\phi(\mathbf X),
u_x(\mathbf X),
u_y(\mathbf X)
\right\}.
\label{eq:single_inclusion_operator}
\end{equation}

\begin{figure}[!htbp]
\centering
\includegraphics[width=0.70\textwidth]{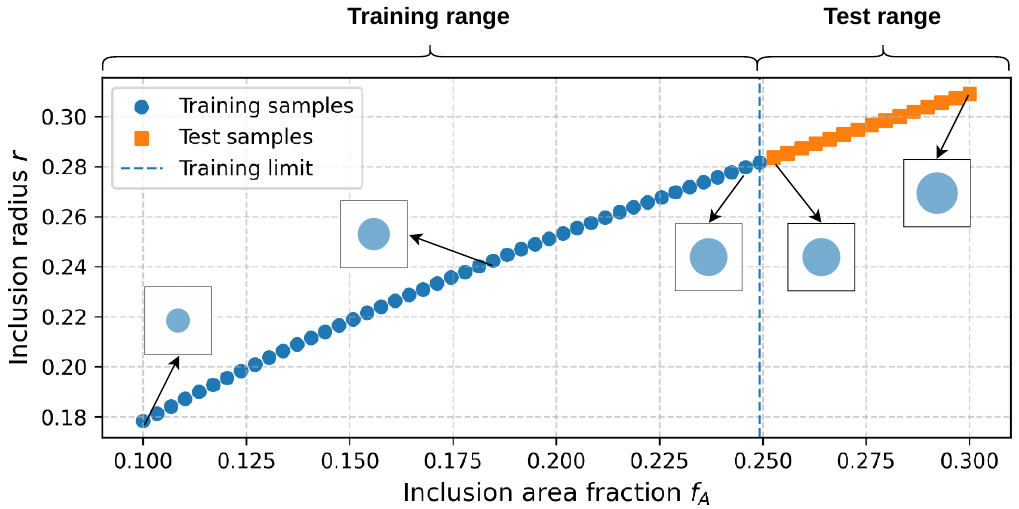}
\caption{Training and test distribution of the scalar control parameter for the single-inclusion area-fraction study.}
\label{fig:area_fraction_control_parameter}
\end{figure}

A total of 60 area-fraction samples are generated in the interval
\begin{equation}
f_A\in[0.10,0.30].
\label{eq:single_inclusion_area_range}
\end{equation}
The first 45 samples are used for training, and the remaining 15 samples are used for testing. Figure~\ref{fig:area_fraction_control_parameter} illustrates the distribution of the scalar control parameter used in the single-inclusion area-fraction study. As shown in Figure~\ref{fig:area_fraction_control_parameter}, the training samples cover the lower range of the area fraction, approximately \(f_A\in[0.10,0.249]\), whereas the test samples cover the larger area fractions, approximately \(f_A\in[0.253,0.30]\). The dashed vertical line marks the upper limit of the training range.

This setup defines an out-of-distribution study with respect to the scalar control parameter \(f_A\). In other words, the test samples do not introduce a new topology, because all samples still contain a single centered circular inclusion. However, the test samples lie outside the training interval of the area fraction. Therefore, this study evaluates the extrapolation capability of the WSNO framework for larger inclusion sizes within the same geometric family.

Two complementary error analyses are conducted for the single-inclusion area-fraction study. First, global relative \(L_2\) errors are computed over the entire computational domain for the magnetic scalar potential \(\phi\), horizontal displacement component \(u_x\), magnetic field \(\mathbb H\), and magnetic induction \(\mathbb B\). These errors quantify the full-field prediction accuracy of the WSNO framework relative to the nonlinear finite-element reference solution.

Second, a regional magnetic analysis is performed by separating the domain into the inclusion and matrix regions. This analysis is important because the magnetic response is strongly phase-dependent due to the contrast in relative permeability. Therefore, a global error alone does not fully indicate whether the model captures the magnetic behavior inside each material phase.

\begin{figure}[!htbp]
\centering
\includegraphics[width=0.6\textwidth]{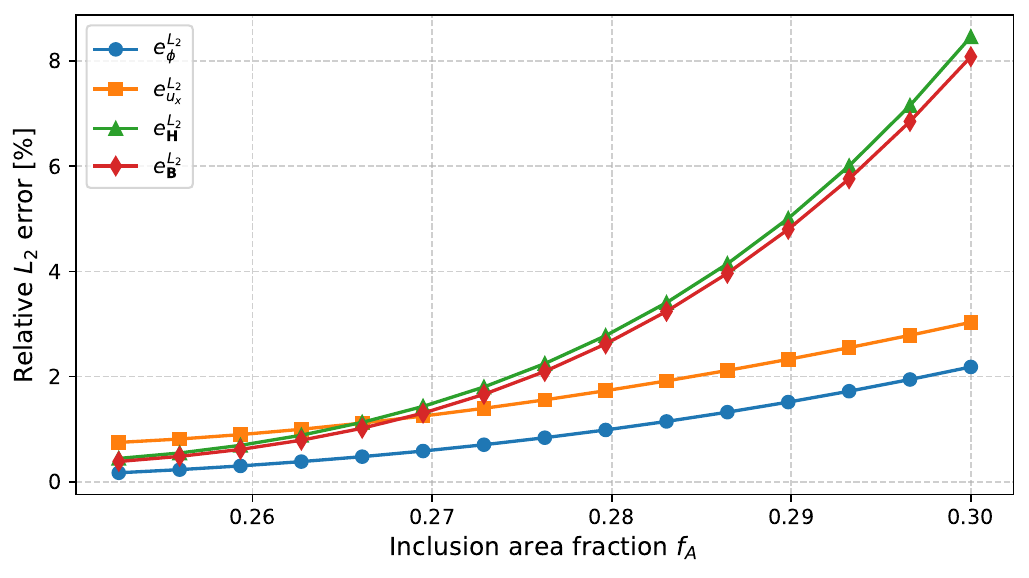}
\caption{Global relative \(L_2\) errors for the out-of-distribution area-fraction test samples.}
\label{fig:area_fraction_global_l2_errors}
\end{figure}

Figure~\ref{fig:area_fraction_global_l2_errors} shows the global relative \(L_2\) errors for the out-of-distribution test samples as a function of the inclusion area fraction \(f_A\). The test samples correspond to larger inclusion area fractions than those used during training and, therefore, evaluate the extrapolation capability of the WSNO model with respect to the scalar control parameter.

The magnetic scalar potential \(\phi\) is predicted with the lowest error over the full test range. The horizontal displacement component \(u_x\), which represents the dominant mechanical response in this boundary-value problem, is also accurately captured. In contrast, the magnetic field \(\mathbb H\) and magnetic induction \(\mathbb B\) exhibit larger errors. This is expected because \(\mathbb H\) is obtained from the spatial gradient of the predicted scalar potential, and \(\mathbb B\) further depends on the heterogeneous permeability field. Therefore, these quantities are more sensitive to local prediction errors near the inclusion-matrix interface. The errors increase as the inclusion area fraction increases. This trend is consistent with the out-of-distribution nature of the test set, since larger area fractions lie farther away from the training interval. Moreover, increasing the inclusion size changes the spatial extent of the high-stiffness and high-permeability phase, leading to stronger perturbations of the coupled magneto-mechanical fields.

\begin{figure}[!htbp]
\centering
\includegraphics[width=0.6\textwidth]{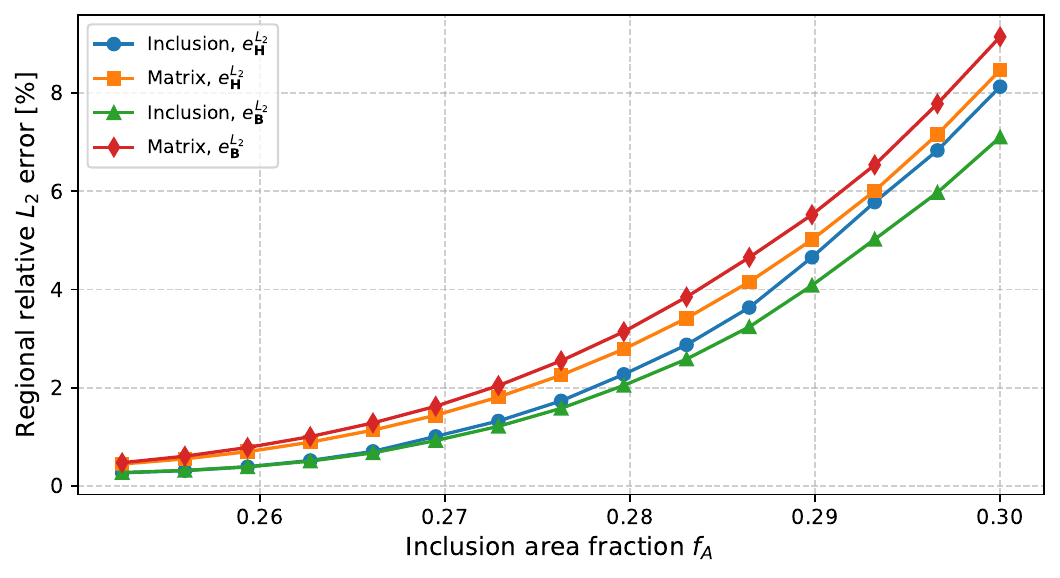}
\caption{Regional relative \(L_2\) errors of the magnetic field \(\mathbb H\) and magnetic induction \(\mathbb B\) for the out-of-distribution area-fraction test samples.}
\label{fig:regional_magnetic_l2_errors_vs_area_fraction}
\end{figure}

Figure~\ref{fig:regional_magnetic_l2_errors_vs_area_fraction} shows the regional relative \(L_2\) errors of the magnetic field \(\mathbb H\) and magnetic induction \(\mathbb B\) as functions of the inclusion area fraction \(f_A\). The errors are reported separately for the inclusion and matrix regions. As the inclusion area fraction increases, the regional errors increase in both phases. This trend is expected because the test samples correspond to larger inclusions than those used during training, and, therefore, represent an extrapolation regime with respect to the scalar control parameter \(f_A\).

It is important to note that the WSNO is trained only on 45 samples with smaller area fractions, while the reported results correspond to 15 out-of-distribution test samples with larger area fractions. Despite this extrapolative setting, the regional magnetic errors remain below \(10\%\) for both \(\mathbb H\) and \(\mathbb B\) in the inclusion and matrix regions. This demonstrates that the proposed model can generalize beyond the training interval and still provide reliable phase-wise magnetic predictions for larger inclusion sizes.

\begin{figure}[!ht]
\centering
\includegraphics[width=0.9\textwidth]{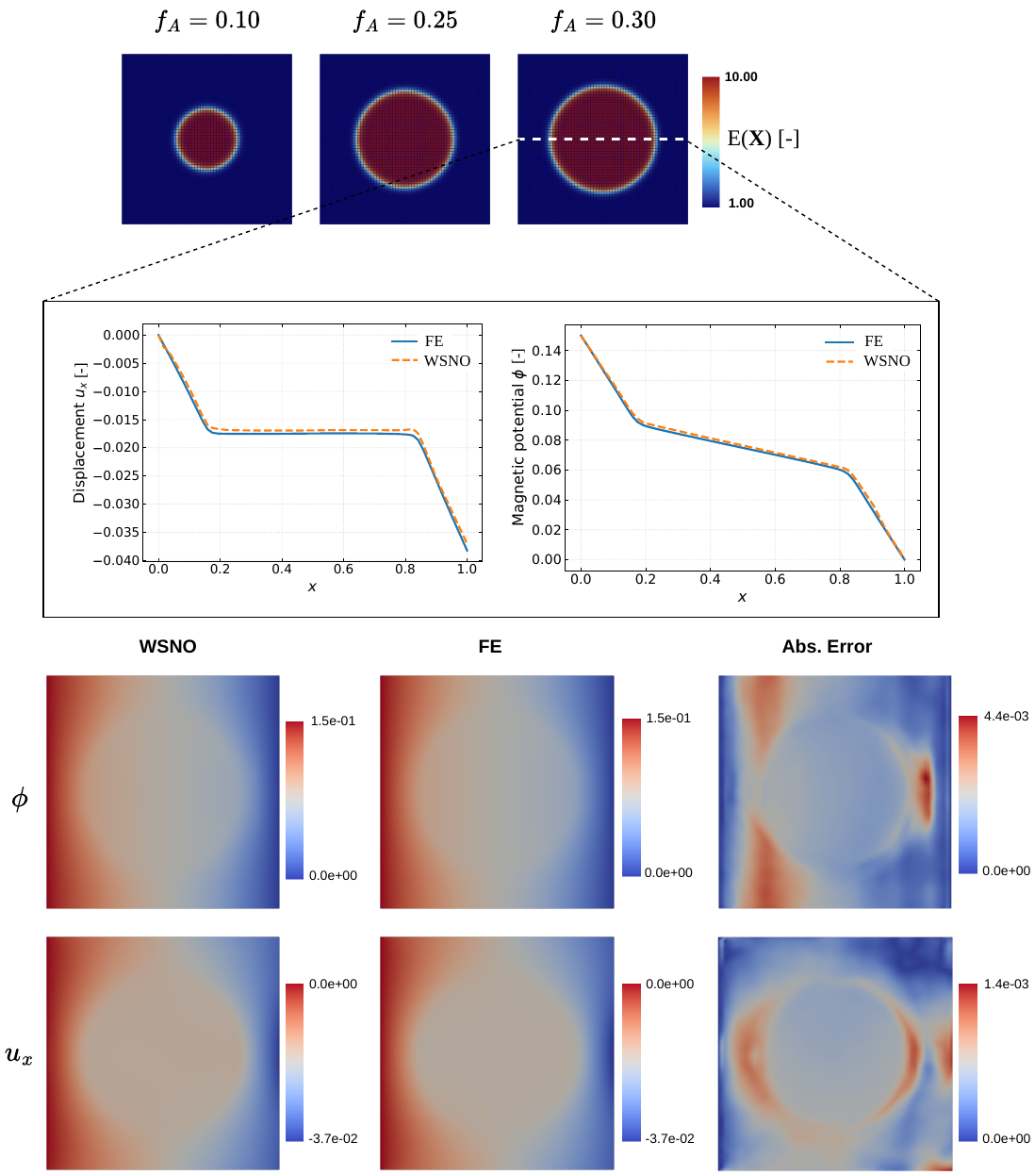}
\caption{Qualitative comparison of the scalar magnetic potential and displacement in x-direction for out-of-distribution single-inclusion test samples with 0.3 area fraction.}
\label{fig:single_inclusion_error}
\end{figure}

Figure~\ref{fig:single_inclusion_error} provides a qualitative comparison of the
WSNO prediction with the finite element reference solution for the single-inclusion problem. The upper part of the figure illustrates the increase in inclusion size for representative area fractions $f_A=0.10$, $0.25$, and $0.30$. For the largest considered area fraction, $f_A=0.30$, a horizontal section through the center of the inclusion is selected to compare the magnetic scalar potential $\phi$ and the horizontal displacement $u_x$. The corresponding line profiles show that the prediction follows the finite element solution closely along the entire section. In particular, the changes in slope associated with the transition between the matrix and inclusion regions are well reproduced, indicating that the learned operator captures the influence of the inclusion size and the material interface on both the magnetic and mechanical responses.
The full-field comparison further confirms this agreement. The WSNO predictions reproduce the overall spatial distributions of both $\phi$ and $u_x$, while the corresponding absolute-error fields remain localized and relatively small compared with the magnitude of the predicted quantities. Larger deviations are mainly observed in the vicinity of the inclusion-matrix interface, where the material properties vary rapidly and the coupled fields exhibit stronger spatial gradients. 

\subsection{Out-of-distribution prediction and neural-initialized
Newton correction}
\label{subsec:ood_nin_study}
Previous studies assess the prediction capability of the proposed neural operators for different microstructures of magneto-active elastomers. In the present study, a more challenging out-of-distribution setting is considered. The network is trained on smooth material distributions generated from random Fourier coefficients. Then, this network is used to evaluate microstructures composed of randomly distributed circular inclusions. Moreover, the neural-initiated Newton (NiN) approach is used to solve the coupled problem in a single load increment~\cite{taghikhani2026neural}. 

The computational domain is defined as
\begin{equation}
\Omega=[0,L]\times[0,L],
\qquad L=1~[-],
\label{eq:nin_domain}
\end{equation}
and is discretized using a structured $64\times64$ quadrilateral mesh.

The matrix and inclusion material properties, as well as the mechanical and magnetic boundary conditions, are adopted from the previous example in Section~\ref{subsec:magneto_mechanical_inclusion} and Figure~\ref{fig:BC_micorstructure}, respectively.

In this study, the same WSNO framework introduced in Section~\ref{subsec:staggered_training} is employed. The magnetic subnetwork is first pretrained for $2000$ epochs with the displacement field set to zero. This stage is followed by two staggered cycles, each consisting of $2000$ mechanical and $2000$ magnetic training epochs. In contrast, the preceding studies employed $20000$ epochs for each training stage. The substantially reduced training effort is intentional, since the objective of the present example is not to obtain a highly accurate direct neural-operator prediction, but rather to investigate whether the neural prediction can provide a sufficiently informative initial state for the subsequent neural-initialized Newton correction.

\subsubsection{Fourier-generated training distributions}

The training microstructures are generated using the Fourier-based parameterization introduced by Taghikhani et al.~\cite{taghikhani2026neural}. In this approach, a reduced set of Fourier coefficients $\boldsymbol{\xi}$ is mapped to a smooth phase-indicator field $w_{\boldsymbol{\xi}}(\mathbf{X}) \in [0,1]$ through a sigmoid projection. The interested reader is referred to reference~\cite{taghikhani2026neural} for the general Fourier representation.

In the present study, the frequency sets in the two spatial directions are chosen as

\begin{equation}
F_x = F_y = \{2,4,6\},
\end{equation}
and the sigmoid sharpness parameter is set to $\beta=20$. The Fourier coefficients are sampled independently from a standard normal distribution.

The resulting phase-indicator field $w_{\boldsymbol{\xi}}(\mathbf{X})$ is used to construct the spatially varying Young's modulus and relative permeability according to the same interpolation relations introduced in Equations~\eqref{eq:young_modulus_interpolation_inclusion} and \eqref{eq:relative_permeability_interpolation_inclusion}, with $w(\mathbf{X})$ replaced by $w_{\boldsymbol{\xi}}(\mathbf{X})$.

\begin{figure}[htbp]
    \centering
    \includegraphics[width=0.75\textwidth]{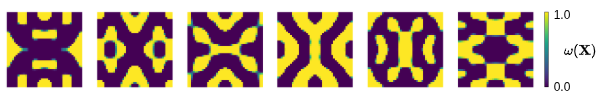}
    \caption{
    Representative Fourier-generated training samples shown in terms of the phase-indicator field $\omega(\mathbf{X})$. The same indicator field is used to construct both the Young's modulus $E(\mathbf{X})$ and the relative magnetic permeability $\mu_r(\mathbf{X})$ through interpolation between the matrix and inclusion properties.
    }
    \label{fig:fourier_training_samples}
\end{figure}

A total of $1000$ independently generated Fourier material distributions are employed for training. Representative samples are shown in Figure~\ref{fig:fourier_training_samples}. In contrast to the circular-inclusion microstructures considered for testing, the Fourier-generated training fields exhibit smooth and irregular material distributions. This deliberate difference in morphology provides the basis for the out-of-distribution evaluation considered in the following section.

\subsubsection{Circular out-of-distribution test microstructures}
Non-overlapping circular inclusions with a fixed radius $r$ are used to generate out-of-distribution test samples. For the $j$-th inclusion centered at $\boldsymbol{c}_j$, the smooth phase-indicator is defined as
\begin{equation}
w_j(\boldsymbol{X})
=
\frac{1}
{
1+
\exp
\left[
\beta
\left(
\left\|
\boldsymbol{X}-\boldsymbol{c}_j
\right\|-r
\right)
\right]
},
\label{eq:nin_circular_indicator}
\end{equation}
where $\boldsymbol{c}_j$ and $r_j$ denote the center and radius of the $j$-th inclusion, respectively, and $\beta$ controls the sharpness of the inclusion-matrix transition. The indicator approaches unity inside the inclusion and zero in the surrounding matrix, while $\omega_j=0.5$ at the nominal interface $\|\boldsymbol{X}-\boldsymbol{c}_j\|=r_j$.

The number of inclusions associated with a prescribed target area fraction $f_A$ is estimated as
\begin{equation}
N_{\mathrm{inc}}
=
\operatorname{round}
\left(
\frac{
f_A A_{\Omega}
}{
\pi r^2
}
\right),
\label{eq:nin_number_inclusions}
\end{equation}
where $A_{\Omega}=L^2~[-]$. Since the number of inclusions must be an integer, the actual geometrical area fraction is
\begin{equation}
f_A^{\mathrm{actual}}
=
\frac{
N_{\mathrm{inc}}\pi r^2
}{
A_{\Omega}
}.
\label{eq:nin_actual_area_fraction}
\end{equation}

\begin{figure}[!ht]
    \centering
    \includegraphics[width=0.90\textwidth]{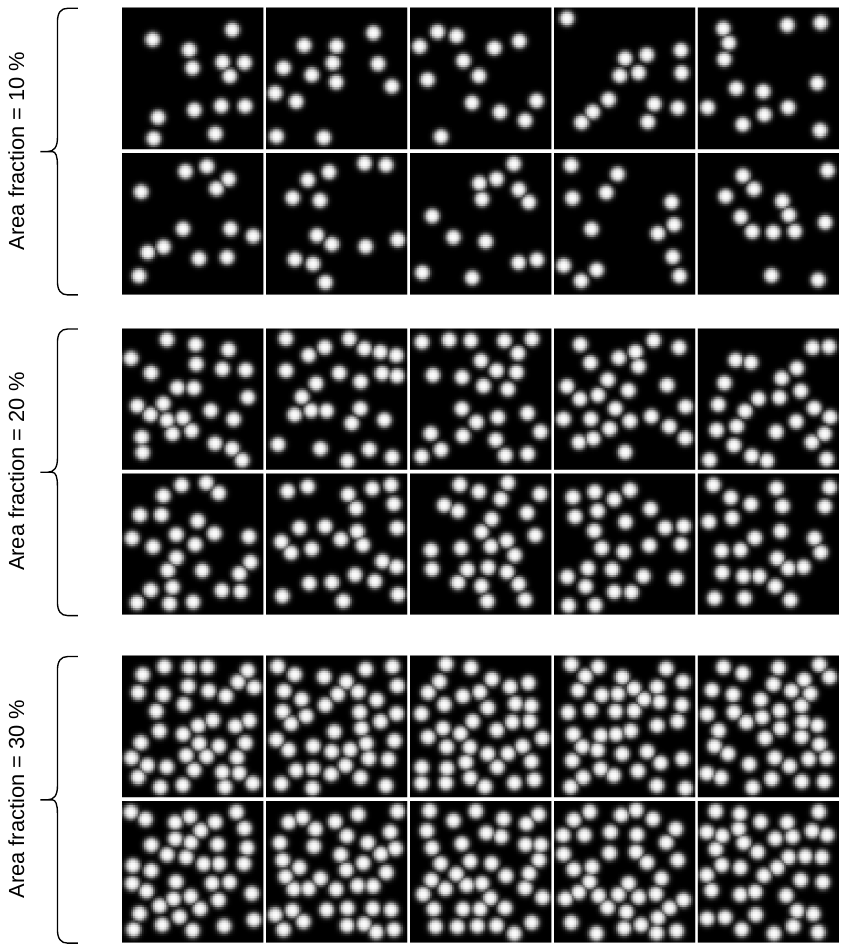}
    \caption{
    Circular out-of-distribution test microstructures considered for inclusion area fractions of $10\%$, $20\%$, and $30\%$.}
    \label{fig:all_circular_ood_samples}
\end{figure}

The target area fractions considered in this study are
\begin{equation}
f_A\in\{0.10,0.20,0.30\},
\label{eq:nin_test_area_fractions}
\end{equation}
with a fixed inclusion radius $r=0.05$. Ten independently generated microstructures are evaluated for each target area fraction, resulting in a total of $30$ out-of-distribution test samples. All test samples are shown in~Figure~\ref{fig:all_circular_ood_samples}. The samples are grouped according to the prescribed inclusion area fraction. For each area fraction, independent realizations are generated by randomly positioning non-overlapping inclusions of fixed radius within the square domain.  

\subsubsection{Neural-initialized Newton correction}

The WSNO prediction is subsequently used to initialize the coupled nonlinear finite element equations. Let
\begin{equation}
\boldsymbol{q}^{(0)}
=
\boldsymbol{q}_{\mathrm{WSNO}}
\label{eq:nin_initial_state}
\end{equation}
denote the initial Newton iterate. At iteration $k$, the coupled residual equations are linearized as
\begin{equation}
\boldsymbol{K}
\left(
\boldsymbol{q}^{(k)}
\right)
\Delta\boldsymbol{q}^{(k)}
=
-
\boldsymbol{R}
\left(
\boldsymbol{q}^{(k)}
\right),
\label{eq:nin_newton_system}
\end{equation}
where
\begin{equation}
\boldsymbol{R}
=
\begin{bmatrix}
\boldsymbol{R}_{\phi}\\
\boldsymbol{R}_{u}
\end{bmatrix}
\label{eq:nin_coupled_residual}
\end{equation}
is the coupled magneto-mechanical residual vector and $\boldsymbol{K}$ is the corresponding tangent operator. The solution is updated according to
\begin{equation}
\boldsymbol{q}^{(k+1)}
=
\boldsymbol{q}^{(k)}
+
\Delta\boldsymbol{q}^{(k)}.
\label{eq:nin_newton_update}
\end{equation}

The Newton correction initialized by the neural prediction is carried out in a single load increment and is denoted by $\boldsymbol{q}_{\mathrm{NiN}}$.

For comparison, an independent nonlinear finite element reference solution is computed from a zero initial state,
\begin{equation}
\boldsymbol{q}_{\mathrm{FE}}^{(0)}
=
\boldsymbol{0},
\label{eq:nin_fe_initial_state}
\end{equation}
using $100$ load increments. This distinction allows the direct neural prediction and the NiN-corrected response to be evaluated against the same converged finite element reference solution.

\subsubsection{Newton convergence and computational efficiency}

The purpose of this section is not to assess the direct predictive capability of the WSNO. Instead, we investigate whether the learned solution can provide an effective initial state for the subsequent nonlinear FE solution. In contrast to the direct WSNO prediction, the neural-initialized Newton (NiN) solution is obtained by solving the same coupled nonlinear finite-element equations as the conventional FE procedure. Therefore, once the prescribed Newton convergence tolerance is reached, the NiN solution is expected to coincide with the converged FE solution up to the specified numerical tolerance. Consequently, the efficiency of NiN is assessed in terms of the number of Newton iterations and the required solution time, rather than by measuring an additional solution error relative to the FE reference.

For each circular out-of-distribution microstructure, the direct WSNO prediction is used as the initial state for the coupled Newton solver. The NiN procedure applies the complete magneto-mechanical loading in a single load increment, whereas the conventional FE reference solution is initialized from the zero state and solved using 100 load increments. Both approaches employ the same nonlinear convergence tolerances and linear-solver settings, as summarized in~\ref{app:hyperparameters}.

Table~\ref{tab:nin_computational_performance} summarizes the computational performance for the three different area fractions considered based on Equation~\eqref{eq:nin_actual_area_fraction}. The reported values correspond to averages over the ten independently generated out-of-distribution microstructures considered at each area fraction. For the conventional FE analysis, the reported Newton-iteration count represents the total number of nonlinear iterations accumulated over all 100 load increments. In contrast, NiN solves the problem in a single load increment.

\begin{table*}[!ht]
\centering
\small
\setlength{\tabcolsep}{10pt}

\caption{Computational performance of the neural-initialized Newton (NiN)
strategy and the conventional nonlinear finite-element (FE) solution
procedure for the out-of-distribution circular-inclusion study.}
\label{tab:nin_computational_performance}

\begin{tabular}{c l l c c c}
\hline
$f_A$
& Method
& Initial state
& \shortstack{Load\\increments}
& \shortstack{Newton\\iterations}
& \shortstack{Mean solution\\time [s]} \\
\hline

10\%
& NiN
& WSNO prediction
& 1
& 4-5
& 2.98 \\

& FE
& Zero state
& 100
& 400
& 261.4 \\
\hline

20\%
& NiN
& WSNO prediction
& 1
& 4-5
& 2.87 \\

& FE
& Zero state
& 100
& 400
& 261.3 \\
\hline

30\%
& NiN
& WSNO prediction
& 1
& 4-5
& 2.47 \\

& FE
& Zero state
& 100
& 400
& 238.9 \\
\hline
\end{tabular}
\end{table*}

Across all investigated area fractions, the NiN procedure converges within approximately four to five Newton iterations, whereas the conventional FE procedure requires approximately 400 Newton iterations in total over the 100 load increments. The corresponding solution times exhibit the same trend. The mean NiN solution time remains below approximately 3~s for all three area fractions, while the conventional FE simulations require approximately 239--261~s. These results demonstrate that the WSNO prediction provides a sufficiently informative initial state to enable convergence of the fully coupled nonlinear problem within a single load increment.
It is worth mentioning that the NiN procedure involves a one-time offline training cost. For the present study, which consists of one magnetic preconditioning stage followed by two staggered training cycles, the total training time was approximately 10.27~h. After training, however, the same operator can be reused for all subsequent microstructure evaluations. Therefore, the computational advantage of the NiN strategy becomes particularly relevant in applications involving repeated solutions of related nonlinear magneto-mechanical boundary-value problems.

\begin{figure}[!ht]
    \centering
    \includegraphics[width=\textwidth]{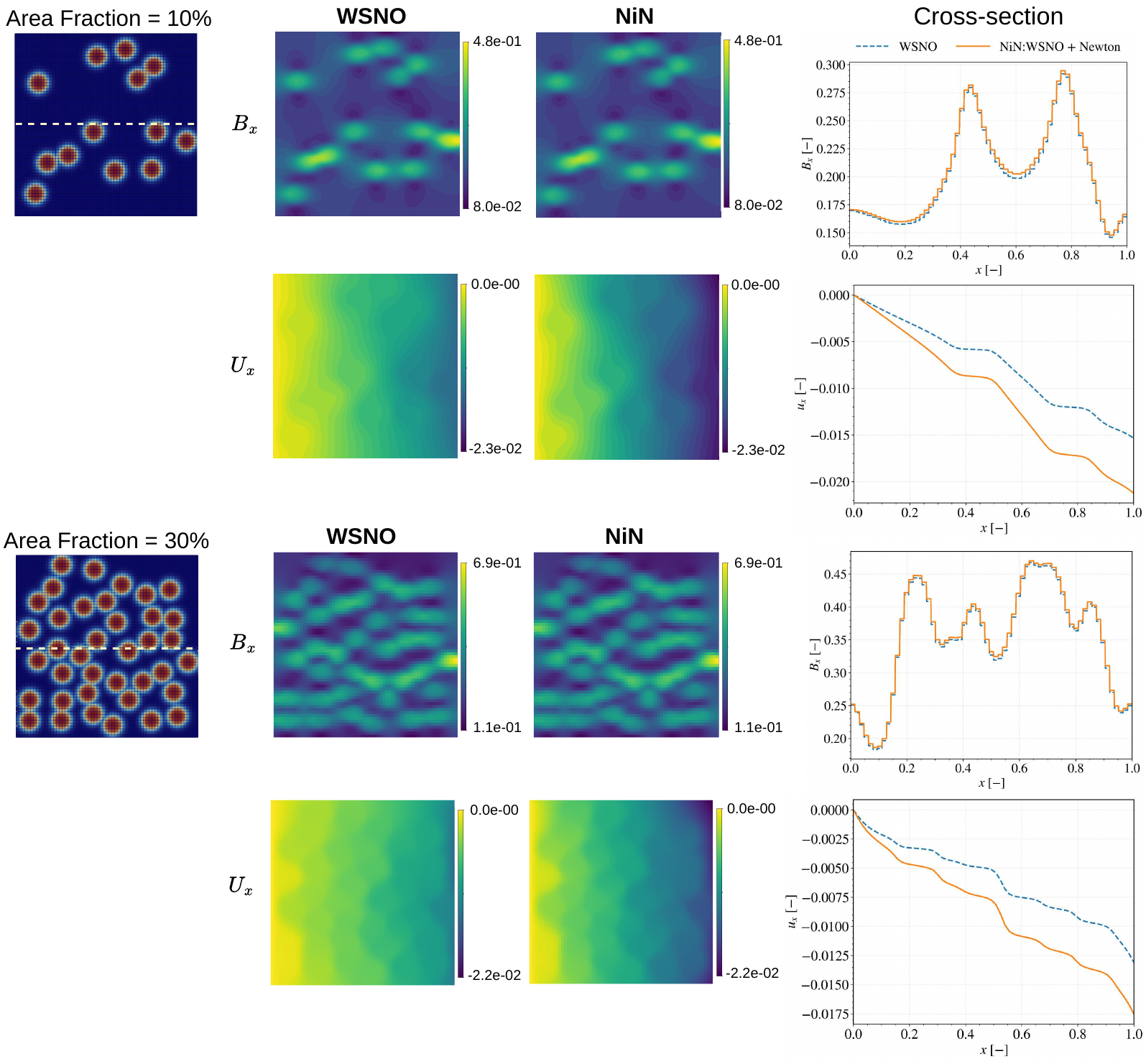}
    \caption{
    Comparison of WSNO and NiN performance on unseen microstructures}
    \label{fig:NiN_micro}
\end{figure}

Figure~\ref{fig:NiN_micro} provides a comparison between the direct WSNO prediction and the NiN-corrected solution for representative out-of-distribution microstructures with area fractions of 10\% and 30\%. The magnetic induction component $\mathbb{B}_x$ shows close agreement between the WSNO and NiN solutions, both in the full-field distributions and along the selected horizontal cross-sections. In contrast, more pronounced corrections are observed for the displacement component $u_x$, indicating that the mechanically induced response under the applied magnetic field is more challenging for the WSNO to predict accurately under the present training procedure. The NiN procedure therefore improves the mechanical response by using the WSNO prediction as the initial guess and subsequently correcting it through the nonlinear Newton iterations.

\subsection{Geometry-parametric prediction of magnetostrictive response of hollow magneto-active cylinders}
\label{subsec:hollow_cylinder}

In this section, the capability of the proposed framework is extended to a three dimensional geometrically problems of magnetostrictive response of the MAEs with the hollow cylinder shape. The geometry of the mentioned cylinder is parameterized independently by the outer radius $R_o$, inner radius $R_i$, and cylinder length $L$. The corresponding geometry-parameter vector is defined as
\begin{equation}
\boldsymbol{\mu}
=
\begin{bmatrix}
R_o & R_i & L
\end{bmatrix}^{\mathrm T}.
\label{eq:hollow_geometry_parameter_vector}
\end{equation}

The geometric parameters are sampled independently within the intervals
\begin{equation}
R_o \in [0.6,1.0],
\qquad
R_i \in [0.15,0.35],
\qquad
L \in [0.6,1.2].
\label{eq:hollow_geometry_ranges}
\end{equation}

Since the inner and outer radii are varied independently, the wall thickness
\begin{equation}
t = R_o-R_i
\label{eq:hollow_wall_thickness}
\end{equation}
also changes throughout the parameter space. Consequently, the present example is more general than a uniform radial scaling of a solid cylinder, since geometries with similar outer dimensions can possess substantially different cavity sizes and wall thicknesses.
The resulting magnetic and mechanical responses therefore depend on three independent geometric parameters.

All constitutive parameters are kept fixed such that the variations in the solution fields arise solely from changes in geometry. The normalized Young's modulus and Poisson's ratio are set to
\begin{equation}
E=1~[-],
\qquad
\nu=0.30,
\end{equation}
while the magneto-mechanical constitutive parameters based on the Equation~\eqref{eq:nonlinear_magnetic_response} are
\begin{equation}
c_1=-4~[-],
\qquad
c_2=2~[-].
\end{equation}

The normalized scalar magnetic potential is prescribed on the bottom and top surfaces of the hollow cylinder according to
\begin{equation}
\phi = 0.50~[-]
\qquad
\text{on } \Gamma_{\phi}^{\mathrm{bottom}},
\label{eq:hollow_phi_bottom}
\end{equation}
and
\begin{equation}
\phi = 0~[-]
\qquad
\text{on } \Gamma_{\phi}^{\mathrm{top}}.
\label{eq:hollow_phi_top}
\end{equation}

For the mechanical problem, all displacement components are constrained at the bottom
surface,
\begin{equation}
u_x=u_y=u_z=0
\qquad
\text{on } \Gamma_{u}^{\mathrm{bottom}},
\label{eq:hollow_bottom_bc}
\end{equation}

whereas only the transverse displacement components are constrained at the top surface,
\begin{equation}
u_x=u_y=0
\qquad
\text{on } \Gamma_{u}^{\mathrm{top}}.
\label{eq:hollow_top_bc}
\end{equation}

Accordingly, the axial displacement $u_z$ remains free on the top surface, allowing the specimen to undergo magnetically induced deformation in the axial direction.

\subsubsection{Fixed-topology mapping of the hollow-cylinder geometry}
\label{subsubsec:hollow_mapping}

In the geometry-parametric operator formulation, all geometry
realizations share the same number and ordering of nodal degrees of freedom. To this end, a hollow cylinder reference is introduced with
\begin{equation}
R_{o,0}=0.9~[-],
\qquad
R_{i,0}=0.25~[-],
\qquad
L_0=1.0~[-].
\label{eq:hollow_reference_geometry}
\end{equation}

The reference geometry is employed only to define a common finite-element topology. For a node with reference coordinates
$\boldsymbol{X}=[X,Y,Z]^{\mathrm T}$, the corresponding reference radial coordinate is
\begin{equation}
r_{\mathrm{ref}}
=
\sqrt{X^2+Y^2}.
\label{eq:hollow_reference_radius}
\end{equation}

The normalized radial position of the node through the wall thickness is defined as
\begin{equation}
\xi_r
=
\frac{
r_{\mathrm{ref}}-R_{i,0}
}{
R_{o,0}-R_{i,0}
}.
\label{eq:hollow_normalized_radial_coordinate}
\end{equation}

Thus, $\xi_r=0$ corresponds to the inner cylindrical surface, whereas $\xi_r=1$ corresponds to the outer surface. For a target geometry characterized by $(R_o,R_i,L)$, the corresponding physical radial coordinate is obtained from
\begin{equation}
r
=
R_i+\xi_r(R_o-R_i).
\label{eq:hollow_radial_mapping}
\end{equation}

The physical nodal coordinates are subsequently computed as
\begin{equation}
x
=
\frac{r}{r_{\mathrm{ref}}}X,
\qquad
y
=
\frac{r}{r_{\mathrm{ref}}}Y,
\qquad
z
=
\frac{L}{L_0}Z.
\label{eq:hollow_coordinate_mapping}
\end{equation}

\begin{figure}[!ht]
    \centering
    \includegraphics[width=\textwidth]
    {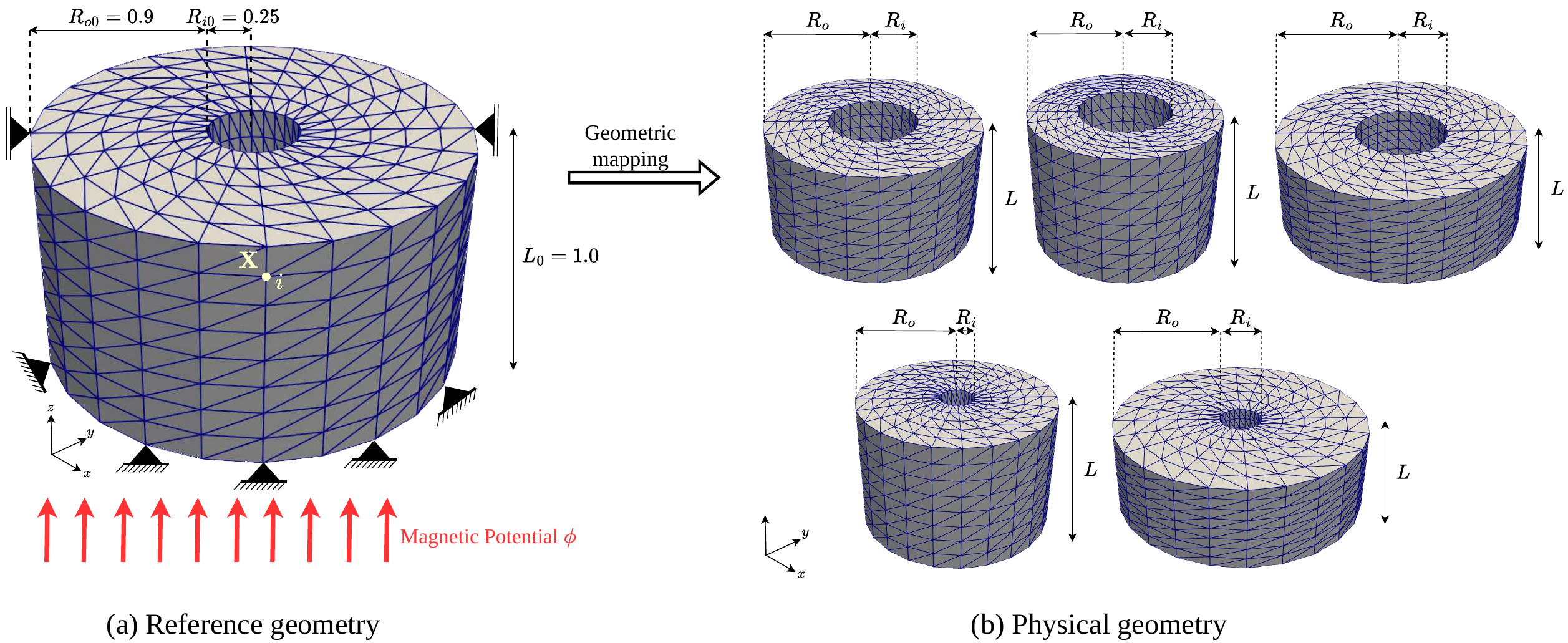}
    \caption{
    Geometry parameterization and sampling strategy for the
    geometry-parametric hollow-cylinder problem.
    (a) Reference hollow-cylinder geometry characterized by
    $R_{o,0}$, $R_{i,0}$, and $L_0$.
    (b) Corresponding physical geometry obtained for a target parameter vector $(R_o,R_i,L)$ while preserving the common finite-element topology.}
    \label{fig:hollow_geometry_mapping_samples}
\end{figure}

This transformation preserves both the circumferential position and the normalized through-thickness position of every node while allowing $R_o$, $R_i$, and $L$ to vary independently. As a consequence, all geometry realizations retain the same node numbering, element connectivity, and degree-of-freedom ordering, whereas only the physical nodal coordinates are modified. The reference-to-physical mapping is illustrated in
Figure~\ref{fig:hollow_geometry_mapping_samples}(a,b). A finite-element mesh of the hollow cylindrical domain is discretized using five radial, 24 circumferential, and eight axial subdivisions. 

It is worth noting that the mapped physical coordinates are used directly in the finite-element evaluation of the element Jacobians, shape-function gradients, numerical quadrature, and assembled magneto-mechanical residuals. Hence, the geometric parameters do not enter the learning problem solely as additional neural-network features; they also modify the physics-informed residual through the geometry-dependent finite-element operators.

\subsubsection{Geometry-parametric neural operator}
\label{subsubsec:hollow_operator}

The unknown nodal state is defined as
\begin{equation}
\boldsymbol{d}_i
=
\begin{bmatrix}
\phi_i &
u_{x,i} &
u_{y,i} &
u_{z,i}
\end{bmatrix}^{\mathrm T},
\label{eq:hollow_nodal_unknowns}
\end{equation}
and the complete solution vector is written as
\begin{equation}
\boldsymbol{y}
=
\begin{bmatrix}
\boldsymbol{d}_1^{\mathrm T} &
\boldsymbol{d}_2^{\mathrm T} &
\cdots &
\boldsymbol{d}_{N_{\mathrm{node}}}^{\mathrm T}
\end{bmatrix}^{\mathrm T}.
\label{eq:hollow_complete_solution}
\end{equation}

Two field-specific Fourier neural operators are employed to represent the magnetic and mechanical solution maps. The magnetic operator receives the constitutive and geometric descriptors
\begin{equation}
\left[
E,\,
c_1,\,
c_2,\,
R_o,\,
R_i,\,
L,\,
R_o-R_i
\right]
\end{equation}
and predicts the scalar magnetic potential,
\begin{equation}
\phi_{\boldsymbol{\theta}_{\phi}}
=
\mathcal{G}_{\phi}
\left(
E,c_1,c_2,R_o,R_i,L,R_o-R_i;
\boldsymbol{\theta}_{\phi}
\right).
\label{eq:hollow_phi_operator}
\end{equation}

The mechanical operator additionally receives the predicted magnetic potential and returns the three displacement components,
\begin{equation}
\boldsymbol{u}_{\boldsymbol{\theta}_{u}}
=
\mathcal{G}_{u}
\left(
E,c_1,c_2,R_o,R_i,L,R_o-R_i,
\phi_{\boldsymbol{\theta}_{\phi}};
\boldsymbol{\theta}_{u}
\right).
\label{eq:hollow_u_operator}
\end{equation}

Accordingly, the learned geometry-to-solution mapping can be summarized as
\begin{equation}
\mathcal{G}:
(R_o,R_i,L)
\longmapsto
\left\{
\phi(\boldsymbol{x}),
\boldsymbol{u}(\boldsymbol{x})
\right\}.
\label{eq:hollow_solution_operator}
\end{equation}

The wall thickness $R_o-R_i$ is explicitly supplied as an additional geometric descriptor, while the same geometry parameters are independently used in the coordinate mapping of
Equation~\eqref{eq:hollow_coordinate_mapping}. This allows the operator to learn the geometry-dependent solution fields while the corresponding weak-form residuals are evaluated on the actual physical geometry.

\subsubsection{Training and test geometries}
\label{subsubsec:hollow_sampling}

A total of 100 geometry realizations are generated by independently sampling $R_o$, $R_i$, and $L$ within the ranges of Equation~\eqref{eq:hollow_geometry_ranges}. The first 75 realizations are employed for physics-informed training. The remaining geometries are
excluded from training, and ten unseen samples are selected for the finite-element-based evaluation reported below.

\begin{figure}[!ht]
    \centering
    \includegraphics[width=\textwidth]
    {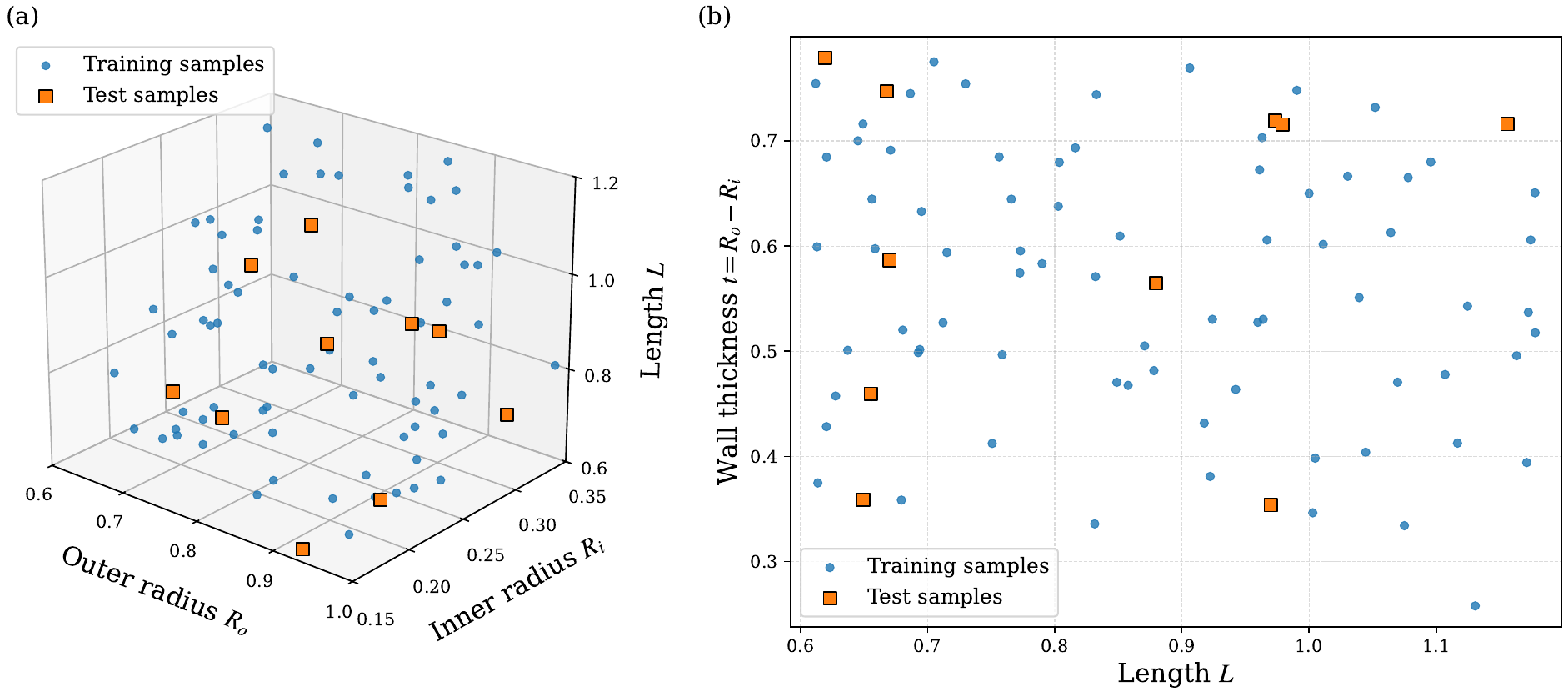}
    \caption{
    (a) Distribution of the training and evaluated test geometries in the
    three-dimensional parameter space $(R_o,R_i,L)$.
    (b) The same samples represented in terms of the cylinder length $L$
    and wall thickness $t=R_o-R_i$.
    }
    \label{fig:hollow_geometry_samples}
\end{figure}

The distribution of the training and evaluated test geometries is shown in Figure~\ref{fig:hollow_geometry_samples}.
Figure~\ref{fig:hollow_geometry_samples}(a) presents the samples in the three-dimensional parameter space $(R_o,R_i,L)$, while
Figure~\ref{fig:hollow_geometry_samples}(b) provides an alternative representation in terms of the wall thickness $t=R_o-R_i$ and cylinder length $L$.

\subsubsection{Prediction accuracy}
\label{subsubsec:hollow_cylinder_accuracy}

After training, nonlinear finite-element analyses are performed independently for the selected unseen hollow-cylinder geometries and are used exclusively as reference solutions for evaluating the neural-operator predictions. No finite-element solution fields are supplied during the physics-informed training procedure.

The relative $L_2$ error of the magnetic scalar potential is defined as
\begin{equation}
e_{\phi}^{L_2}
=
\frac{
\left\|
\phi_{\mathrm{WSNO}}
-
\phi_{\mathrm{FE}}
\right\|_2
}{
\left\|
\phi_{\mathrm{FE}}
\right\|_2
}
\times 100 ,
\label{eq:hollow_phi_error}
\end{equation}
whereas the relative error of the complete three-dimensional displacement field is evaluated as
\begin{equation}
e_{\boldsymbol{u}}^{L_2}
=
\frac{
\left\|
\boldsymbol{u}_{\mathrm{WSNO}}
-
\boldsymbol{u}_{\mathrm{FE}}
\right\|_2
}{
\left\|
\boldsymbol{u}_{\mathrm{FE}}
\right\|_2
}
\times 100 .
\label{eq:hollow_u_error}
\end{equation}

The quantitative errors obtained for the ten unseen geometries are summarized in Table~\ref{tab:hollow_cylinder_errors}. The magnetic scalar potential is predicted with consistently high accuracy over the investigated parameter space. The corresponding mean relative error of the complete three-dimensional displacement field is approximately \(0.76\%\). For eight of the ten evaluated geometries, the displacement error remains below approximately \(1.00\%\). The largest displacement error, approximately \(1.84\%\), is observed for sample~76.
\begin{table}[htbp]
    \centering
    \caption{Relative $L_2$ errors of the magnetic scalar potential and
    three-dimensional displacement field for the ten evaluated unseen
    hollow-cylinder geometries.}
    \label{tab:hollow_cylinder_errors}
    \begin{tabular}{ccccccc}
        \hline
        Sample
        & $R_o$
        & $R_i$
        & $L$
        & $t = R_o - R_i$
        & $e_{\phi}^{L_2}$ [\%]
        & $e_{\mathbf{u}}^{L_2}$ [\%] \\
        \hline
        75 & 0.610 & 0.251 & 0.649 & 0.359 & 0.031 & 0.572 \\
        76 & 0.869 & 0.153 & 1.156 & 0.716 & 0.126 & 1.844 \\
        77 & 0.968 & 0.249 & 0.973 & 0.719 & 0.016 & 0.591 \\
        78 & 0.931 & 0.344 & 0.670 & 0.586 & 0.042 & 0.442 \\
        79 & 0.954 & 0.207 & 0.668 & 0.747 & 0.045 & 0.478 \\
        80 & 0.864 & 0.300 & 0.880 & 0.564 & 0.040 & 0.540 \\
        81 & 0.698 & 0.239 & 0.655 & 0.460 & 0.019 & 0.401 \\
        82 & 0.907 & 0.192 & 0.979 & 0.716 & 0.019 & 0.594 \\
        83 & 0.685 & 0.331 & 0.970 & 0.354 & 0.072 & 1.352 \\
        84 & 0.933 & 0.153 & 0.619 & 0.779 & 0.091 & 0.761 \\
        \hline
        Mean & -- & -- & -- & -- & 0.050 & 0.758 \\
        \hline
    \end{tabular}
\end{table}

\begin{figure}[!ht]
    \centering
    \includegraphics[width=0.64\textwidth]
    {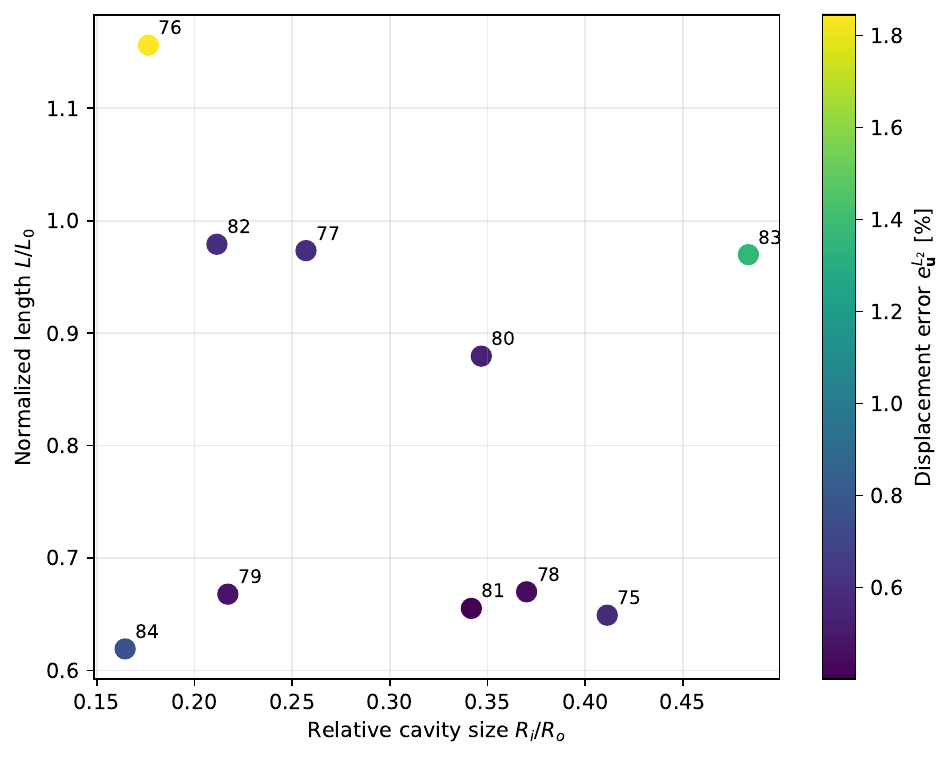}
    \caption{
    Geometry-dependent distribution of the relative displacement error for the ten evaluated unseen hollow-cylinder geometries.}
    \label{fig:hollow_geometry_error_map}
\end{figure}

To further assess the influence of geometric configuration on the mechanical prediction accuracy, two dimensionless geometric parameters are considered. Figure~\ref{fig:hollow_geometry_error_map} presents the relative displacement error as a function of the relative cavity size, $R_i/R_o$, and the cylinder length normalized by the reference length, $L/L_0$. The color of each marker represents the corresponding relative $L_2$ error of the complete three-dimensional displacement field. The error distribution shows that the mechanical prediction accuracy depends on the combined geometric configuration rather than on a single geometric parameter. Sample~76 which shows the largest displacement error, also has the largest normalized length, with $L/L_o \approx 1.16$. In contrast, sample~83 exhibits the second-largest displacement error, despite having a considerably smaller normalized length and the largest relative cavity size with $R_i/R_o \approx 0.48$. These observations suggest that both the cavity size and the axial dimension influence the difficulty of the geometry-dependent mechanical prediction.

\begin{figure}[!ht]
    \centering
    \includegraphics[width=0.95\textwidth]
    {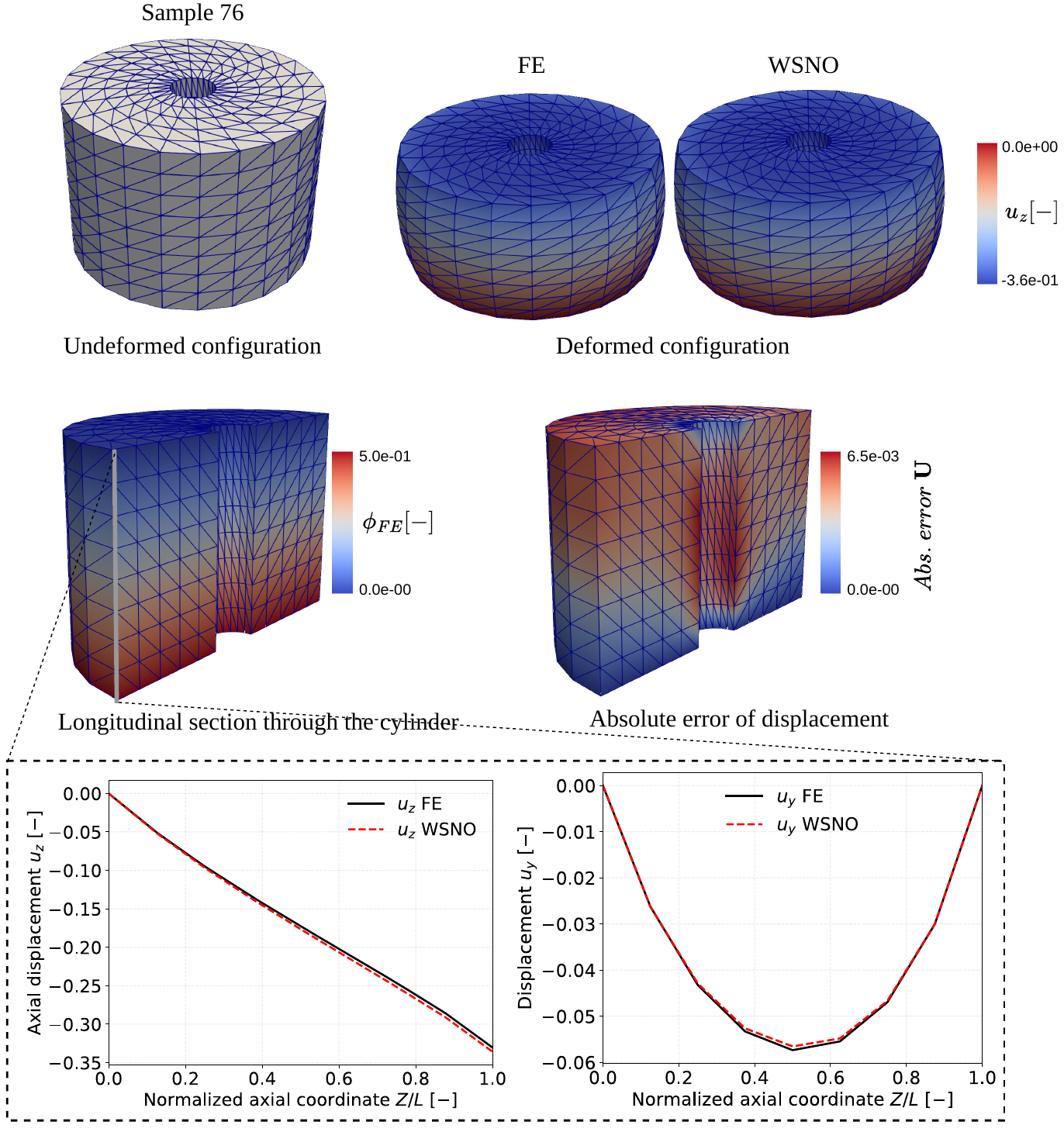}
    \caption{ Qualitative comparison of the magnetostrictive response for sample~76.The undeformed configuration, FE and WSNO deformed configurations, a longitudinal section through the hollow cylinder, and the corresponding displacement profiles $u_z(z/L)$ and $u_y(z/L)$ are shown.}
    \label{fig:hollow_magnetostriction}
\end{figure}

To assess the predicted magnetostrictive response, sample~76 as the most challenging case is selected. Figure~\ref{fig:hollow_magnetostriction} compares the undeformed configuration with the deformed geometries obtained from the nonlinear finite-element solution and the WSNO prediction. The overall deformation pattern predicted by the WSNO closely follows the finite-element reference, including the magnetically induced axial deformation and the accompanying change in the hollow-cylinder geometry. A longitudinal section through the specimen is additionally considered to provide a clearer view of the deformation through the cylinder wall. The displacements $u_z$ and $u_y$ are further extracted along a representative axial line and plotted as a function of the normalized axial coordinate $z/L$. As shown in Figure~\ref{fig:hollow_magnetostriction}, the WSNO prediction closely follows the finite-element solution over the entire cylinder length, including the region of largest axial displacement near the free end. 

To further study of the resolution generalization capability of the proposed WSNO framework, a zero-shot super-resolution (ZSSR) study is evaluated for the geometry-parametric hollow cylinder problem. In this study, the network parameters obtained at the training resolution are kept fixed. Then, the trained operator is assessed directly on a finer spatial discretizations, without any additional training. Therefore, the study examines whether the learned operator can reproduce the coupled magneto-mechanical solution to a resolution not encountered during training. 
The original discretization is defined by $(n_r,n_\theta,n_z)=(5,24,8)$, corresponding to an FNO grid of
$(9,24,6)$. For the ZSSR evaluation, the radial and circumferential resolutions are increased to
$(n_r,n_\theta,n_z)=(10,48,8)$, resulting in an FNO grid of
$(9,48,11)$, while the axial resolution is retained. The same trained network parameters are used at both resolutions. For each of the ten unseen geometries, an independent finite element solution is computed on the corresponding spatial discretization and used as the reference for assessing the WSNO prediction.

\begin{figure}[htbp]
    \centering
    \includegraphics[width=0.9\textwidth]{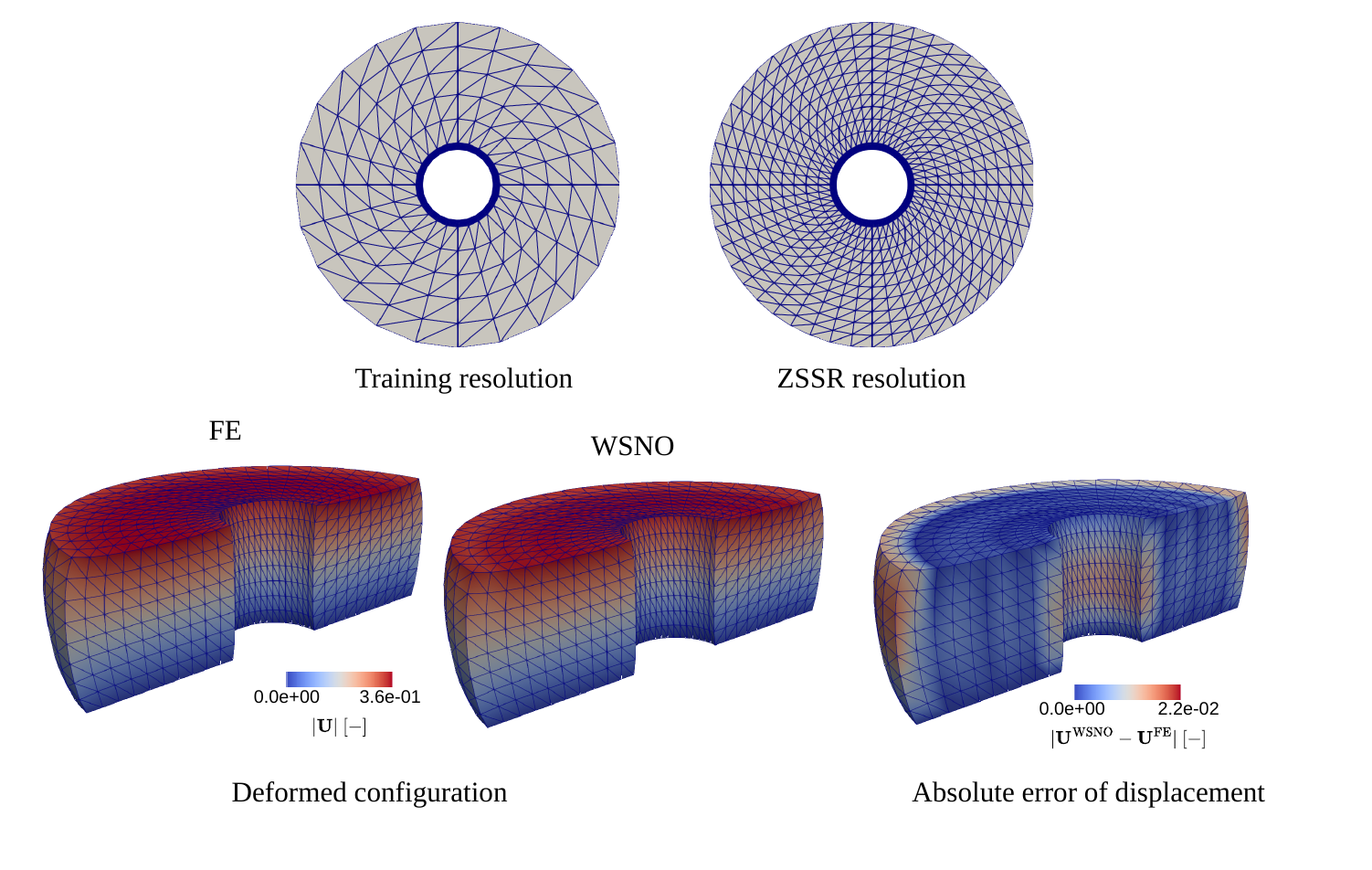}
    \caption{Zero-shot super-resolution comparison for sample~77, showing the
    training and ZSSR discretizations, the FE and WSNO deformed configurations,
    and the corresponding absolute displacement error.}
    \label{fig:zssr_sample77}
\end{figure}

The result demonstrate that the trained WSNO can be evaluated directly on the finer discretization without retraining. As expected, the prediction errors increase when moving from the training resolution to the finer ZSSR resolution, reflecting
the greater spatial detail that must be represented by the same set of learned parameters. Nevertheless, the model preserves the overall coupled magneto-mechanical response and provides physically consistent predictions on the refined grid. This indicates that the learned operator is not restricted to the specific discretization employed during training, but retains a degree of mesh-resolution transferability. Table~\ref{tab:hollow_cylinder_zssr} summarizes the zero-shot super-resolution results for the ten unseen geometries. At the refined resolution, the mean relative \(L_2\) errors of the magnetic scalar potential and the three-dimensional displacement field are \(0.60\%\) and \(2.73\%\), respectively. Despite the increased spatial resolution and the absence of any additional training or fine-tuning, the WSNO maintains low prediction errors across all evaluated geometries. Moreover, the mean WSNO inference time is approximately \(0.128\,\mathrm{s}\), compared with approximately \(1387\,\mathrm{s}\) for the corresponding finite element solution, highlighting the substantial computational efficiency of the learned operator. Figure~\ref{fig:zssr_sample77} illustrates the zero-shot super-resolution performance for sample~77 by comparing the coarse training discretization with the refined ZSSR grid. The FE and WSNO deformed configurations show close agreement, while the absolute displacement-error field remains small relative to the overall displacement magnitude.

\begin{table}[htbp]
    \centering
    \caption{Zero-shot super-resolution prediction errors and computational
    times for the ten unseen hollow-cylinder geometries.}
    \label{tab:hollow_cylinder_zssr}
    \begin{tabular}{ccccc}
        \hline
        Sample
        & $e_{\phi}^{L_2}$ [\%]
        & $e_{\mathbf{u}}^{L_2}$ [\%]
        & $t_{\mathrm{WSNO}}$ [s]
        & $t_{\mathrm{FE}}$ [s] \\
        \hline
        75 & 0.638 & 2.200 & 0.129 & 1386.2 \\
        76 & 0.586 & 3.652 & 0.127 & 1383.0 \\
        77 & 0.579 & 3.177 & 0.128 & 1379.6 \\
        78 & 0.620 & 2.462 & 0.128 & 1391.4 \\
        79 & 0.601 & 2.407 & 0.128 & 1385.1 \\
        80 & 0.597 & 2.908 & 0.127 & 1384.5 \\
        81 & 0.625 & 2.280 & 0.127 & 1390.5 \\
        82 & 0.580 & 3.126 & 0.128 & 1381.9 \\
        83 & 0.612 & 2.874 & 0.128 & 1385.4 \\
        84 & 0.606 & 2.224 & 0.128 & 1405.2 \\
        \hline
        Mean & \textbf{0.604} & \textbf{2.731}
             & \textbf{0.128} & \textbf{1387.3} \\
        \hline
    \end{tabular}
\end{table}

\FloatBarrier
\section{Conclusion}
\label{sec:conclusion}

In this study, a data-free physics-informed operator-learning framework has been developed for the finite strain coupled magneto-mechanical response of magneto-active elastomers. The governing magnetostatic and mechanical equations are incorporated through finite-element weak-form residuals, allowing the neural operators to be trained without reference finite-element solution fields. The proposed WSNO framework employs separate magnetic and mechanical operators and alternates their optimization while retaining the coupled constitutive interaction between the magnetic potential and displacement fields. This staggered procedure is introduced to overcome the poor convergence observed for direct monolithic optimization of the coupled physics-informed loss.

The performance of the framework is first demonstrated on heterogeneous two-dimensional microstructures containing randomly distributed magnetic inclusions. The WSNO accurately reproduced the magnetic scalar potential and displacement fields for unseen in-distribution samples, while the larger errors observed in the magnetic-field components are mainly associated with their recovery through spatial differentiation of the predicted scalar potential. These results confirmed that the proposed staggered formulation can capture the coupled response of heterogeneous magneto-active composites while preserving the interaction between the magnetic and mechanical fields.

The generalization capability of the learned operator is further investigated using a single-inclusion geometry-parametric problem in which the inclusion area fraction of the test cases exceeded the training interval. Despite this extrapolative setting, the magnetic potential and dominant displacement component remained accurately predicted, with mean relative $L_2$ errors of approximately $0.97\%$ and $1.68\%$, respectively. The magnetic field and magnetic induction exhibited moderately higher errors because of their increased sensitivity to the inclusion-matrix interface and local field gradients. Nevertheless, the regional magnetic errors remained below $10\%$ in both the inclusion and matrix phases, indicating that the learned operator retained meaningful phase-wise predictive capability outside the training interval.

A substantially more demanding out-of-distribution setting is subsequently considered by training the WSNO on smooth Fourier-generated material distributions and evaluating it on microstructures composed of circular inclusions. In this case, the magnetic quantities retained moderate predictive accuracy, whereas the direct mechanical predictions showed poor generalization. This result highlights an important limitation of direct operator extrapolation when the morphology of the test samples differs significantly from that encountered during training. At the same time, the WSNO predictions are shown to provide effective initial states for a subsequent neural-initialized Newton correction. The corrected solutions converged closely to the nonlinear finite-element reference solutions across all investigated inclusion area fractions, demonstrating the value of combining operator learning with a physics-based nonlinear solver.

Finally, the proposed WSNO framework is extended to a three-dimensional geometry-parametric problem describing the magnetostrictive response of hollow magneto-active cylinders. The geometry is parameterized independently by the outer radius $R_o$, inner radius $R_i$, and cylinder length $L$, while a common finite-element topology is preserved through a fixed-topology mapping from a reference configuration to each physical geometry. In this way, the learned operator captures the dependence of the coupled magnetic and mechanical solution fields on variations in cavity size, wall thickness, and specimen length. For ten unseen geometries, the mean relative $L_2$ errors are approximately $0.11\%$ for the magnetic scalar potential and $1.36\%$ for the three-dimensional displacement field. These results demonstrate that the proposed data-free physics-informed formulation can accurately capture geometry-dependent magnetostrictive responses without requiring labeled finite-element solution data.

Overall, the presented results show that weak-form physics-informed operator learning provides a promising framework for rapid prediction of coupled magneto-mechanical boundary-value problems across families of heterogeneous material distributions and geometries. The staggered optimization strategy improves the robustness of training for the coupled magnetic and mechanical fields, while the neural-initialized Newton approach offers a practical mechanism for recovering high-accuracy solutions in cases where direct out-of-distribution generalization is insufficient.

Future study should focus on extending the framework to more complex three-dimensional heterogeneous microstructures, varying loading and boundary conditions, and constitutive models incorporating anisotropy, magnetic saturation, and rate-dependent behavior. In addition, integrating the present framework into multiscale and optimization frameworks represents a promising direction for accelerating repeated nonlinear magneto-mechanical simulations in the design and analysis of magneto-active materials.

\section*{Author contributions}

Alireza Yazdandousthamedani: Conceptualization, Methodology, Software, Validation, Investigation, Writing - Original Draft, Writing - Review \& Editing. Ahmad Moeineddin: Conceptualization, Methodology, Writing - Review \& Editing. Reza Najian Asl: Software, Validation, Writing - Review \& Editing. Shahed Rezaei: Conceptualization, Methodology, Validation, Investigation, Supervision, Writing - Review \& Editing. Michael Kaliske: Funding acquisition, Supervision, Writing - Review \& Editing.

\section*{Acknowledgments}
The ﬁnancial support within the research project No. 380321452 (GRK 2430, Sub-Project 6) by the Deutsche Forschungsgemeinschaft (DFG, German Research Foundation) is gratefully acknowledged.

The authors gratefully acknowledge the computing time made available to them on the high-performance computer at the NHR Center of TU Dresden. This center is jointly supported by the Federal Ministry of Research, Technology and Space of Germany and the state governments participating in the NHR (www.nhr-verein.de/unsere-partner).

\section*{Data Availability Statement}
The data that support the findings of this study are openly available in
folax at \url{https://github.com/Neural-Mechanics-Lab/folax}.

\appendix

\section{Hyperparameters and numerical settings}
\label{app:hyperparameters}

The principal neural-operator architectures and training hyperparameters employed in the numerical examples are summarized in Table~\ref{tab:operator_hyperparameters}. The corresponding problem-specific discretization, material, and parametric settings are reported in Table~\ref{tab:problem_hyperparameters}. The studies in Sections~\ref{sec:results} employ the FNO-based WSNO framework.


\begin{table*}[!ht]
\centering
\caption{Neural-operator architectures and training hyperparameters employed
in the numerical examples.}
\label{tab:operator_hyperparameters}

\renewcommand{\arraystretch}{1.10}
\setlength{\tabcolsep}{3pt}
\small

\resizebox{\textwidth}{!}{%
\begin{tabular}{lccccc}
\hline
\textbf{Hyperparameter}
&
\textbf{Sec.~\ref{subsec:magneto_mechanical_inclusion}}
&
\textbf{Sec.~\ref{subsubsec:phase_contrast}}
&
\textbf{Sec.~\ref{subsec:area_fraction_single_inclusion}}
&
\textbf{Sec.~\ref{subsec:ood_nin_study}}
&
\textbf{Sec.~\ref{subsec:hollow_cylinder}}
\\
\hline

Operator architecture
& WSNO--FNO
& WSNO--FNO
& WSNO--FNO
& WSNO--FNO
& WSNO-- 3D~FNO
\\

Magnetic FNO modes
& $12\times12$
& $12\times12$
& $12\times12$
& $12\times12$
& $9\times12\times6$
\\

Magnetic FNO width
& 64 & 64 & 64 & 64 & 64
\\

Magnetic FNO depth
& 6
& 6
& 6
& 4
& 6
\\

Mechanical FNO modes
& $12\times12$
& $16\times16$
& $12\times12$
& $12\times12$
& $9\times12\times6$
\\

Mechanical FNO width
& 64
& 96
& 64
& 64
& 64
\\

Mechanical FNO depth
& 6
& 6
& 6
& 4
& 6
\\

Projection dimension
& 256
& 256
& 256
& 256
& 256
\\

Magnetic output scale
& $10^{-2}$
& $10^{-2}$
& $10^{-2}$
& $10^{-2}$
& $10^{-2}$
\\

Mechanical output scale
& $10^{-3}$
& $10^{-3}$
& $10^{-3}$
& $10^{-3}$
& $10^{-3}$
\\

Optimizer
& Adam
& Adam
& Adam
& Adam
& Adam
\\

Magnetic pretraining epochs
& 2000
& 2000
& 2000
& 2000
& 2000
\\

Number of staggered cycles
& 3
& 3
& 3
& 2
& 3
\\

Mechanical epochs per cycle
& 20000
& 20000
& 15000
& 2000
& 10000
\\

Magnetic epochs per cycle
& 20000
& 20000
& 10000
& 2000
& 10000
\\

Coupled fine-tuning epochs
& 0
& 0
& 0
& 0
& 1000
\\

Mechanical learning rate
& $2\times10^{-5}\rightarrow10^{-6}$
& $2\times10^{-5}\rightarrow10^{-6}$
& $2\times10^{-5}\rightarrow10^{-6}$
& $2\times10^{-5}\rightarrow10^{-6}$
& $2\times10^{-5}\rightarrow10^{-6}$
\\

Magnetic learning rate
& $10^{-4}\rightarrow10^{-5}$
& $10^{-4}\rightarrow10^{-5}$
& $10^{-4}\rightarrow10^{-5}$
& $10^{-4}\rightarrow10^{-5}$
& $10^{-4}\rightarrow10^{-5}$
\\

Batch size
& 5
& 5
& 5
& 5
& 5
\\

\hline
\end{tabular}%
}

\end{table*}


\begin{table*}[ht]
\centering
\caption{Problem-specific discretization, material, and parametric settings employed in the numerical examples.}
\label{tab:problem_hyperparameters}

\renewcommand{\arraystretch}{1.10}
\setlength{\tabcolsep}{3pt}
\small

\resizebox{\textwidth}{!}{%
\begin{tabular}{lccccc}
\hline
\textbf{Parameter}
&
\textbf{Sec.~\ref{subsec:magneto_mechanical_inclusion}}
&
\textbf{Sec.~\ref{subsubsec:phase_contrast}}
&
\textbf{Sec.~\ref{subsec:area_fraction_single_inclusion}}
&
\textbf{Sec.~\ref{subsec:ood_nin_study}}
&
\textbf{Sec.~\ref{subsec:hollow_cylinder}}
\\
\hline

Spatial dimension
& 2D
& 2D
& 2D
& 2D
& 3D
\\

Spatial discretization
& $42\times42$
& $42\times42$
& $64\times64$
& $64\times64$
& tetrahedral
\\

Training microstructures/geometries
& 50
& 50 / 100
& 45
& 1000
& 75
\\

Test microstructures/geometries
& 10
& 10
& 15
& 10 per area fraction
& 10
\\

Training morphology
& circular inclusions
& circular inclusions
& single circular inclusion
& Fourier-generated fields
& hollow cylinder \\
\\

Testing morphology
& circular inclusions
& circular inclusions
& single circular inclusion
& circular inclusions
& hollow cylinder \\
\\

Number of inclusions
& 15
& 15
& 1
& variable
& --
\\

Inclusion radius
& $[0.05,0.10]$
& $[0.05,0.10]$
& determined by $f_A$
& 0.05
& --
\\

Inclusion gap
& 0.015
& 0.015
& --
& 0.005
& --
\\

Interface sharpness $\beta$
& 100
& 100
& 100
& 100
& --
\\

Fourier-control sharpness $\beta$
& --
& --
& --
& 20
& --
\\

Fourier frequencies $F_x$
& --
& --
& --
& $\{2,4,6\}$
& --
\\

Fourier frequencies $F_y$
& --
& --
& --
& $\{2,4,6\}$
& --
\\

Matrix Young's modulus $E_m$
& 1
& 1
& 1
& 1
& 1
\\

Inclusion Young's modulus $E_i$
& 10
& 10
& 10
& 10
& --
\\

Matrix relative permeability $\mu_{r,m}$
& 1
& 1
& 1
& 1
& --
\\

Inclusion relative permeability $\mu_{r,i}$
& 5
& variable
& 10
& 5
& --
\\

Training $\mu_{r,i}$
& 5
& $\{2,5,10\}$
& 10
& 5
& --
\\

Evaluation $\mu_{r,i}$
& 5
& $\{4,5,8\}$
& 10
& 5
& --
\\

Area-fraction range $f_A$
& --
& --
& $[0.10,0.30]$
& $\{0.10,0.20,0.30\}$
& --
\\

Outer-radius range $R_o$
& --
& --
& --
& --
& $[0.6,1.0]$
\\

Inner-radius range $R_i$
& --
& --
& --
& --
& $[0.15,0.35]$
\\

Length range $L$
& --
& --
& --
& --
& $[0.6,1.2]$
\\

Reference outer radius $R_{o,0}$
& --
& --
& --
& --
& 0.9
\\

Reference inner radius $R_{i,0}$
& --
& --
& --
& --
& 0.25
\\

Reference length $L_0$
& --
& --
& --
& --
& 1.0
\\

Radial mesh divisions
& --
& --
& --
& --
& 5
\\

Circumferential mesh divisions
& --
& --
& --
& --
& 24
\\

Axial mesh divisions
& --
& --
& --
& --
& 8
\\

Poisson's ratio $\nu$
& 0.3
& 0.3
& 0.3
& 0.3
& 0.3
\\

$c_1$
& --
& --
& --
& --
& $-4.0$
\\

$c_2$
& --
& --
& --
& --
& $2.0$
\\

\hline
\end{tabular}%
}

\end{table*}


For the phase-contrast study in Section~\ref{subsubsec:phase_contrast}, each distinct training microstructure is combined with the three inclusion relative-permeability values
\begin{equation}
    \mu_{r,i}^{\mathrm{train}}\in\{2,5,10\}.
\end{equation}
Consequently, the cases employing 50 and 100 distinct training microstructures correspond to 150 and 300 microstructure--contrast realizations, respectively. The same 10 unseen microstructures are retained for evaluation in both cases.


In the out-of-distribution study of Section~\ref{subsec:ood_nin_study}, the WSNO is trained on 1000 smooth material distributions generated from random Fourier coefficients using the frequency sets
\begin{equation}
    F_x=F_y=\{2,4,6\}.
\end{equation}
The trained operator is subsequently evaluated on circular-inclusion microstructures with target area fractions of $10\%$, $20\%$, and $30\%$, with 10 independently generated realizations considered for each area fraction. The direct WSNO prediction is subsequently employed as the initial state for the neural-initialized Newton correction.


\begin{table}[!htbp]
\centering
\caption{Nonlinear solver settings employed for the neural-initialized Newton correction and the finite element reference solution in Section~\ref{subsec:ood_nin_study}.}
\label{tab:nin_solver_settings}

\renewcommand{\arraystretch}{1.15}
\setlength{\tabcolsep}{8pt}
\small

\begin{tabular}{lcc}
\hline
\textbf{Solver parameter}
& \textbf{NiN}
& \textbf{FE reference}
\\
\hline

Initial state
& WSNO prediction
& zero state
\\

Load increments
& 1
& 100
\\

Maximum Newton iterations
& 12
& 12
\\

Relative tolerance
& $10^{-6}$
& $10^{-6}$
\\

Absolute tolerance
& $10^{-6}$
& $10^{-6}$
\\

Linear solver
& JAX-direct
& JAX-direct
\\

Linear tolerance
& $10^{-6}$
& $10^{-6}$
\\

Preconditioner
& ILU
& ILU
\\

\hline
\end{tabular}

\end{table}

\FloatBarrier
\section{Normalization of physical quantities}
\label{app:normalization}
The numerical examples are formulated in normalized form so that the
reported results are independent of a particular choice of physical
length and stiffness scales. Let $L_0$ and $E_0$ denote arbitrary
characteristic length and mechanical stiffness scales, respectively.
The normalized spatial coordinates, displacement, elastic modulus,
stress, and energy density are defined as
\begin{equation}
    \bar{\mathbf X}
    =
    \frac{\mathbf X}{L_0},
    \qquad
    \bar{\mathbf u}
    =
    \frac{\mathbf u}{L_0},
    \qquad
    \bar E
    =
    \frac{E}{E_0},
    \qquad
    \bar{\mathbf P}
    =
    \frac{\mathbf P}{E_0},
    \qquad
    \bar{\Psi}
    =
    \frac{\Psi}{E_0}.
    \label{eq:normalization_mechanical}
\end{equation}

For the magnetic quantities, let $\mathbb{H}_0$ denote the characteristic
magnetic-field scale. The normalized magnetic field is defined as
\begin{equation}
    \bar{\mathbb H}
    =
    \frac{\mathbb{H}}{\mathbb{H}_0}.
\end{equation}
Using the magnetic-energy scale $\mu_0^{\mathrm{phys}} \mathbb{H}_0^2$ relative
to the mechanical-energy scale $E_0$, the normalized free-space permeability can be written as
\begin{equation}
    \bar{\mu}_0
    =
    \frac{
        \mu_0^{\mathrm{phys}} \mathbb{H}_0^2
    }{E_0},
    \label{eq:normalized_mu0}
\end{equation}
where $\mu_0^{\mathrm{phys}}$ denotes the physical permeability of free space. In the present numerical formulation, the magnetic scale is selected such that
\begin{equation}
    \bar{\mu}_0=1.
\end{equation}
Consequently,
\begin{equation}
    \mathbb{H}_0
    =
    \sqrt{
        \frac{E_0}{\mu_0^{\mathrm{phys}}}
    }.
    \label{eq:H_reference}
\end{equation}

Since the magnetic field is related to the scalar potential through
$\mathbb H=-\nabla_{\mathbf X}\phi$, the corresponding scalar-potential
scale is
\begin{equation}
    \phi_0
    =
    L_0 \mathbb{H}_0
    =
    L_0
    \sqrt{
        \frac{E_0}{\mu_0^{\mathrm{phys}}}
    },
\end{equation}
and therefore
\begin{equation}
    \bar{\phi}
    =
    \frac{\phi}{\phi_0}.
\end{equation}

The characteristic magnetic-induction scale is chosen as
\begin{equation}
    \mathbb{B}_0
    =
    \mu_0^{\mathrm{phys}} \mathbb{H}_0
    =
    \sqrt{
        \mu_0^{\mathrm{phys}} E_0
    },
\end{equation}
such that
\begin{equation}
    \bar{\mathbb B}
    =
    \frac{\mathbb B}{\mathbb{B}_0}.
\end{equation}

The deformation-dependent magnetic contribution introduced in Equation~\eqref{eq:nonlinear_magnetic_response} is expressed as
\begin{equation}
    \Psi_{\mathrm{coup}}
    =
    c_1 I_6 + c_2 I_7,
\end{equation}
where both invariants $I_6$ and $I_7$ are quadratic in the magnetic
field. Using
\begin{equation}
    \mathbb H = \mathbb{H}_0 \bar{\mathbb H},
\end{equation}
the corresponding invariants scale as
\begin{equation}
    I_6 = \mathbb{H}_0^2 \bar I_6,
    \qquad
    I_7 = \mathbb{H}_0^2 \bar I_7.
\end{equation}
Since the free-energy density is normalized by $E_0$, the normalized
coupling energy becomes
\begin{equation}
    \bar{\Psi}_{\mathrm{coup}}
    =
    \bar c_1 \bar I_6
    +
    \bar c_2 \bar I_7,
\end{equation}
with the dimensionless constitutive parameters
\begin{equation}
    \bar c_1
    =
    \frac{c_1^{\mathrm{phys}} \mathbb{H}_0^2}{E_0},
    \qquad
    \bar c_2
    =
    \frac{c_2^{\mathrm{phys}} \mathbb{H}_0^2}{E_0}.
    \label{eq:normalized_c1_c2}
\end{equation}
Using the normalization condition
\begin{equation}
    \bar{\mu}_0
    =
    \frac{\mu_0^{\mathrm{phys}} \mathbb{H}_0^2}{E_0}
    =1,
\end{equation}
the normalized coupling parameters can equivalently be written as
\begin{equation}
    \bar c_1
    =
    \frac{c_1^{\mathrm{phys}}}{\mu_0^{\mathrm{phys}}},
    \qquad
    \bar c_2
    =
    \frac{c_2^{\mathrm{phys}}}{\mu_0^{\mathrm{phys}}}.
\end{equation}
Thus, dimensional parameters of the form
$c_i^{\mathrm{phys}}=\alpha_i\mu_0^{\mathrm{phys}}$
correspond directly to the normalized coefficients
$\bar c_i=\alpha_i$.

Accordingly, once $L_0$ and $E_0$ are prescribed, all dimensional quantities can be recovered from the normalized results through
Accordingly, once $L_0$ and $E_0$ are prescribed, all dimensional
quantities can be recovered from the normalized results through
\begin{equation}
\begin{aligned}
    \mathbf{X} &= L_0 \bar{\mathbf{X}},
    &\qquad
    \mathbf{u} &= L_0 \bar{\mathbf{u}},
    &\qquad
    E &= E_0 \bar{E}, \\
    \mathbf{P} &= E_0 \bar{\mathbf{P}},
    &
    \Psi &= E_0 \bar{\Psi},
    &
    \phi &= \phi_0 \bar{\phi}, \\
    \mathbb{H} &= \mathbb{H}_0 \bar{\mathbb{H}},
    &
    \mathbb{B} &= \mathbb{B}_0 \bar{\mathbb{B}},
    &
    c_i^{\mathrm{phys}}
    &= \mu_0^{\mathrm{phys}} \bar{c}_i,
    \qquad i=1,2.
\end{aligned}
\label{eq:dimensional_recovery}
\end{equation}

Thus, the normalized numerical results can be interpreted for different
physical length and stiffness scales without modifying the computational
formulation.




\FloatBarrier
\bibliographystyle{elsarticle-num}

\bibliography{Ref}

\end{document}